\documentclass[twocolumn]{aastex631} 

\usepackage{CJK}
\usepackage{longtable}
\usepackage{url}
\usepackage{amsmath}

\begin{document}
\begin{CJK*}{UTF8}{gbsn}

\title{A dense dark matter core of the subhalo in the strong lensing system JVAS B1938+666}

\correspondingauthor{Lei Lei (雷磊)}
\email{leilei@pmo.ac.cn, astro\_lei@outlook.com}

\author[0000-0003-4631-1915]{Lei Lei (雷磊)}
\affiliation{Key Laboratory of Dark Matter and Space Astronomy, Purple Mountain Observatory, Chinese Academy of Sciences, Nanjing 210023, China}
\affiliation{School of Astronomy and Space Science, University of Science and Technology of China, Hefei 230026, China}

\author[0000-0003-1215-6443]{Yi-Ying Wang (王艺颖)}
\affiliation{Key Laboratory of Dark Matter and Space Astronomy, Purple Mountain Observatory, Chinese Academy of Sciences, Nanjing 210023, China}

\author[0000-0001-7540-9335]{Qiao Li (李巧)}
\affiliation{Key Laboratory of Dark Matter and Space Astronomy, Purple Mountain Observatory, Chinese Academy of Sciences, Nanjing 210023, China}
\affiliation{School of Astronomy and Space Science, University of Science and Technology of China, Hefei 230026, China}

\author[0009-0005-6876-7376]{Jiang Dong (董江)}
\affiliation{Key Laboratory of Dark Matter and Space Astronomy, Purple Mountain Observatory, Chinese Academy of Sciences, Nanjing 210023, China}
\affiliation{School of Astronomy and Space Science, University of Science and Technology of China, Hefei 230026, China}

\author[0009-0004-9366-1947]{Ze-Fan Wang (王泽凡)}
\affiliation{Key Laboratory of Dark Matter and Space Astronomy, Purple Mountain Observatory, Chinese Academy of Sciences, Nanjing 210023, China}
\affiliation{School of Astronomy and Space Science, University of Science and Technology of China, Hefei 230026, China}

\author{Wei-Long Lin (林炜龙)}
\affiliation{Key Laboratory of Dark Matter and Space Astronomy, Purple Mountain Observatory, Chinese Academy of Sciences, Nanjing 210023, China}
\affiliation{School of Astronomy and Space Science, University of Science and Technology of China, Hefei 230026, China}

\author[0000-0002-9063-698X]{Yi-Ping Shu (舒轶平)}
\affiliation{Key Laboratory of Dark Matter and Space Astronomy, Purple Mountain Observatory, Chinese Academy of Sciences, Nanjing 210023, China}
\affiliation{School of Astronomy and Space Science, University of Science and Technology of China, Hefei 230026, China}

\author[0000-0003-4988-9296]{Xiao-Yue Cao (曹潇月)}
\affiliation{School of Astronomy and Space Science, University of Chinese Academy of Sciences, Beĳing 100049, China}
\affiliation{National Astronomical Observatories, Chinese Academy of Sciences, 20A Datun Road, Chaoyang District, Beĳing 100012, China}

\author[0000-0002-5421-3138]{Da-Neng Yang (杨大能)}
\affiliation{Key Laboratory of Dark Matter and Space Astronomy, Purple Mountain Observatory, Chinese Academy of Sciences, Nanjing 210023, China}

\author[0000-0002-8966-6911]{Yi-Zhong Fan (范一中)}
\affiliation{Key Laboratory of Dark Matter and Space Astronomy, Purple Mountain Observatory, Chinese Academy of Sciences, Nanjing 210023, China}
\affiliation{School of Astronomy and Space Science, University of Science and Technology of China, Hefei 230026, China}



\begin{abstract}
The nature of dark matter remains unknown, motivating the study of fuzzy/wave dark matter (FDM/$\psi$DM) and self-interacting dark matter (SIDM) as alternative frameworks to address small-scale discrepancies in halo profiles inferred from observations. This study presents a non-parametric reconstruction of the mass distribution of the previously-found, dark subhalo in the strong-lensing system JVAS B1938+666. 
Compared with the standard Navarro-Frenk-White (NFW) profile, both SIDM and $\psi$DM ($m_{\psi}=1.32^{+0.22}_{-0.31}\times 10^{-22} \, \rm eV$) provide significantly better fits to the resulting density profile.
Moreover, the SIDM model is favored over $\psi$DM with a Bayes factor of 14.44. The reconstructed density profile features a characteristic kiloparsec-scale core ($r_c \approx 0.5 \, \rm kpc$) with central density $\rho_c \approx 2.5\times 10^{7}\, \rm M_{\odot} \, kpc^{-3} $, exhibiting remarkable consistency with the core-halo mass scaling relations observed in Local Group dwarf spheroidals. These findings offer insights that may help address the core-cusp discrepancy in $\Lambda$CDM substructure predictions.
\end{abstract}


\section{Introduction} \label{sec:intro}

Dark matter comprises $\sim 84\%$ of the cosmic matter budget and $\sim 25\%$ of the total energy density \citep{2016A&A...594A..13P,2024arXiv240319236K}, yet its fundamental nature remains one of the most pressing mysteries in modern physics. Over nine decades of astronomical observations and laboratory experiments \citep[see reviews by][]{2019A&ARv..27....2S,2024arXiv240601705C} have constrained but not conclusively identified dark matter's particle properties, assuming that dark matter consists of  elementary particles. The theoretical landscape spans an extraordinary 36 orders of magnitude in mass, ranging from ultra-light scalar fields ($m_\psi \sim 10^{-24}$ eV) predicted by string compactifications to TeV-scale supersymmetric particles \citep{1978PhRvL..40..223W,2021ARA&A..59..247H,2024arXiv240317697O,2013IJMPA..2830042K}. This immense parameter space reflects diverse production mechanisms in the early Universe, from quantum fluctuations during inflation to thermal freeze-out processes \citep{2010ARA&A..48..495F,2011AdAst2011E...8G}.

The cold dark matter (CDM) paradigm, anchored by its success in explaining cosmic microwave background anisotropies \citep{2012AnP...524..507F} and large-scale structure formation \citep{1993ARA&A..31..689O}, encounters persistent challenges on galactic scales. Three key discrepancies emerge: (1) The core-cusp problem, where N-body simulations predict Navarro-Frenk-White (NFW) cusps \citep{1996ApJ...462..563N} contrary to observed flat density cores in dwarf galaxies \citep{2017ARA&A..55..343B}; (2) The missing satellites problem, revealing an order-of-magnitude discrepancy between predicted CDM subhalos and observed Milky Way satellites \citep{2020MNRAS.495...58S}; and (3) The too-big-to-fail problem, where the most massive simulated subhalos exceed the kinematic measurements of bright satellites \citep{2017ARA&A..55..343B}. While direct detection experiments continue probing weakly interacting massive particles (WIMPs) \citep[e.g.,][]{2022PhRvL.129i1802F,2024arXiv240711737F}, alternative dark matter candidates addressing these small-scale challenges have gained wide attention.

Two particularly compelling solutions emerge from distinct physical regimes: (1) Self-interacting dark matter (SIDM) introduces particle collisions that redistribute energy within halos, leading to core formation or, in some cases, core collapse \citep{2000PhRvL..84.3760S,2018PhR...730....1T,Yang:2025xsp}, while (2) fuzzy/wave dark matter (FDM/$\psi$DM) generates quantum pressure through ultra-light bosons ($m_\psi \sim 10^{-22}$ eV) that suppress small-scale power \citep{2000PhRvL..85.1158H,2021ARA&A..59..247H}. These models predict markedly different density profiles for $10^7$-$10^9 M_\odot$ subhalos.

Low-mass substructures in galaxy-galaxy strong gravitational lensing systems provide powerful probes of dark matter physics \citep{2019ApJ...872...11H,2023arXiv230814640S,2018A&A...615A.102W}.By leveraging purely gravitational effects — as substructures perturb the morphology of lensed arcs — these systems are able to constrain dark matter properties without relying on baryonic dynamics. This gravitational lensing approach offers particular advantages for high-redshift studies, where traditional kinematic methods become observationally prohibitive \citep{2023arXiv230611781V,2024SSRv..220...87S}.
Among the identified substructures in the strong lensing systems, three substructures have masses ranging from $10^8$  to $10^{10} \, \rm M_{\odot}$ \citep{2010MNRAS.407..225V,2010MNRAS.408.1969V,2012Natur.481..341V,2024MNRAS.52710480N,2025arXiv250802776S}. The lightest substructure, with a mass $1.9\times 10^8\,\rm M_{\odot}$ \citep{2012Natur.481..341V,2012MNRAS.424.2800L,2014MNRAS.442.3598V}, was identified in the system JVAS B1938+666.

The gravitational strong lens system JVAS B1938+666 was discovered by the Jodrell Bank-Very Large Array (VLA) survey \citep{1997MNRAS.289..450K} and subsequently identified through Hubble Space Telescope (HST) optical observations \citep{1998MNRAS.295L..41K}. This system features a background source at redshift $z_{\mathrm{s}} = 2.059$ \citep{2011ApJ...730..108R} lensed by a primary galaxy at redshift $z_{\mathrm{lens}} = 0.881$ \citep{2000AJ....119.1078T}. Analysis of Keck II high-resolution data by \citet{2012Natur.481..341V} and \citet{2012MNRAS.424.2800L} revealed a low-mass dark subhalo ($M_{sub}= 1.9\pm 0.1 \times10^{8}\, \rm M_{\odot}$) within this system using a pixelated-lens modelling technique. The inner density slope of this subhalo was constrained to $\gamma < 1.6$ with the generalized NFW (gNFW) density profile models \citep{2014MNRAS.442.3598V}. Subsequently, using HST data, \citet{2022MNRAS.516..336S} estimated a steeper density slope of $\gamma = 1.96^{+0.12}_{-0.12}$ for the perturber. Further analysis with Keck II data by \citet{2025A&A...699A.222D} constrained the power-law slope to $\gamma_{\mathrm{pl}} = 2.42 \pm 0.19$. 
The low surface brightness of this subhalo makes it challenging to determine whether it is associated with the main lens (a dark subhalo) or a perturber resides along the line-of-sight. \citet{2025arXiv250507944T} identified it as a foreground halo at $z = 0.13 \pm 0.07$. However, \citet{2022MNRAS.515.4391S} found the redshift of the line-of-sight halo to be $z = 1.42^{+0.1}_{-0.15}$. In this work, we assumed that the substructure is a subhalo in the lens plane. To date, only Keck observatory \citep{2012Natur.481..341V,2012MNRAS.424.2800L} and the radio Karl G. Jansky Very Large Array (VLA) \citep{2020MNRAS.495.2387S} have high-spatial resolution observations of the strong lens system JVAS B1938+666. 

This work presents a novel probe of dark matter physics through the lensing system JVAS B1938+666. We implement a non-parametric reconstruction of the subhalo mass distribution, in order to be agnostic to specific density profile parametrizations. Our approach enables direct comparison among three competing models: (i) the collisionless NFW profile predicted by CDM, (ii) the SIDM profile, and (iii) the solitonic $\psi$DM wave solution. Section~\ref{sec:data} details our data analysis pipeline. Section~\ref{sec:result} compares density profiles of SIDM, $\psi$DM and CDM, while Section~\ref{sec:dis} discusses implications for dark matter microphysics and the $\Lambda$CDM paradigm.

We adopt the \cite{2020A&A...641A...6P} cosmology throughout: $(\Omega_{\rm m}, \Omega_\Lambda, H_0) = (0.3111, 0.6889, 67.66$ km s$^{-1}$ Mpc$^{-1})$ with spatial flatness. 

\section{Data and Analysis}
\label{sec:data}

\subsection{Data Reduction}

We analysed archival near-infrared observations of JVAS B1938+666 obtained with the Keck II telescope's NIRC2 instrument coupled with the laser guide star adaptive optics system. The dataset, originally presented in \cite{2012Natur.481..341V} and \cite{2012MNRAS.424.2800L}, consists of Kp-band exposures acquired during June 2010 (Program ID: U085N2L, accessible via Keck Observatory Archive \href{https://koa.ipac.caltech.edu}{KOA}). The point-spread function (PSF) of the NIRC2 instrument is taken from the Keck PSF reconstruction (PSF-R) project Science Verification in 2020 (Program ID: E339), which was released on \href{https://koa.ipac.caltech.edu/cgi-bin/KOA/nph-KOArel}{KOA}. 

Out of 93 images in the Kp-band, we selected 66 frames with 180 seconds of exposure time for each frame to exclude the low quality images with low signal-to-noise ratio or significant background contamination. Using the Python library \emph{Astropy} \citep{2013A&A...558A..33A,2018AJ....156..123A}, we extracted the images of JVAS B1938+666 with $150 \times 150$ pixels. Following the data reduction procedures of \cite{2012Natur.481..341V}, we applied a sub-pixel shift technique to align the 66 science data frames onto a unified coordinate system after the standard Keck data reduction process. Before stacking, we clipped the bad pixels with values outside $3\, \sigma$ in each frame. We also clipped bad pixels in each pixel array, which was composed of 66 frames in the same position after alignment. We stacked the 66 frames as the final image. 

\begin{figure*}[ht!]
\centering
 \includegraphics[width=2\columnwidth]{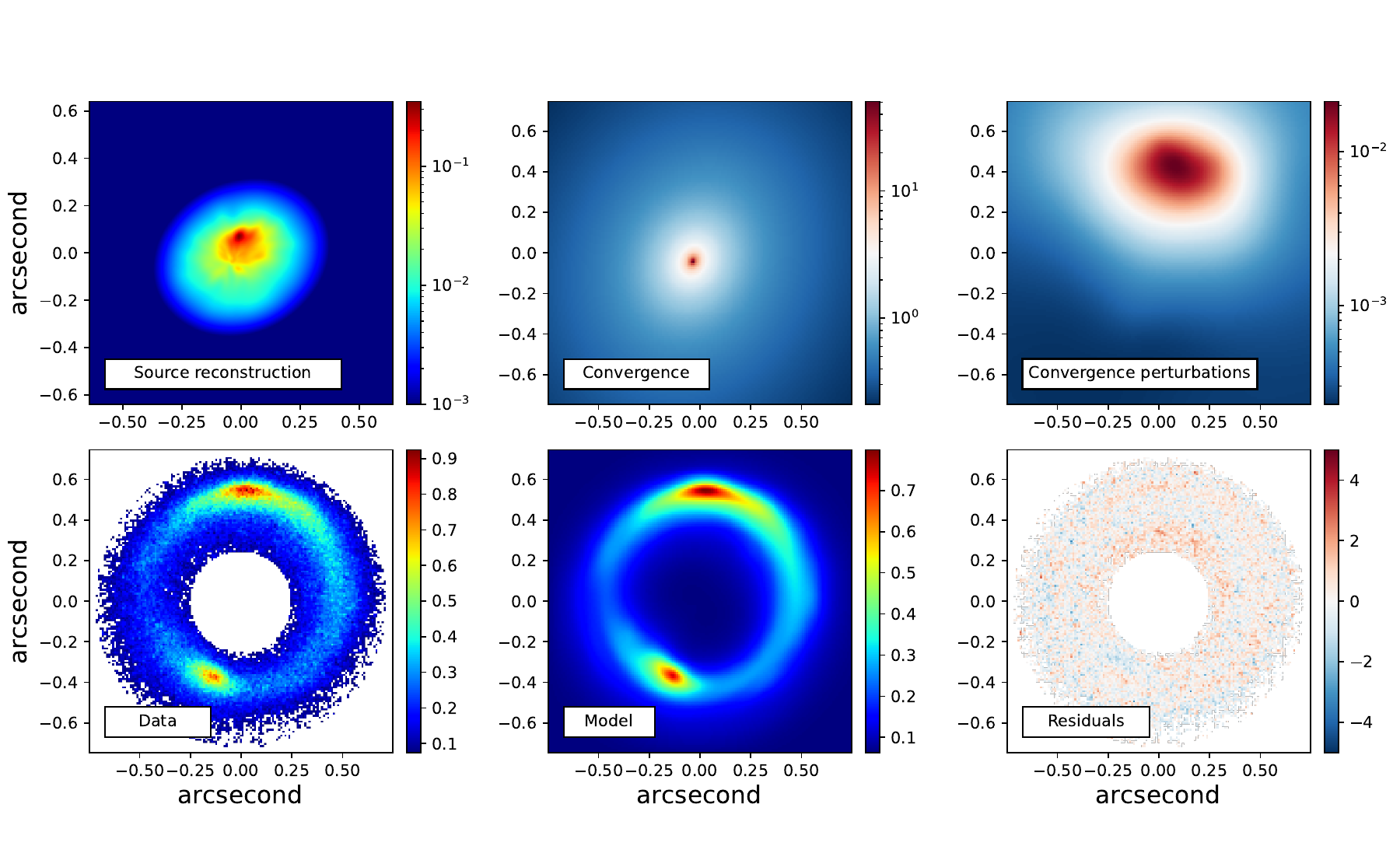}
 \caption{ The detection of the subhalo in JVAS B1938+666 Keck data. The upper panel displays the optimal reconstruction of the source's surface brightness (top left), the lensing convergence (top centre), and the added perturbation in convergence due to the subhalo (top right). The bottom panel presents the Keck/NIRC2 Kp-band data of JVAS B1938+666 in units of $\rm e^- /s $ (bottom left), the best-fit model including PSF convolution (bottom centre), and the normalized residuals between the observed data and the best-fit model (bottom right). The x and y axes of the images are calibrated in arcseconds. }
 \label{fig:subhalo}
\end{figure*}

We then subtracted the median background from a $12 \times 12$ pixel square area, located in the corner region outside the lensing image. Similarly, as \cite{2012Natur.481..341V}, we subtracted the lens light of the central main galaxy by fitting two S\'ersic models \citep{1963BAAA....6...41S}. After subtracting the lens light, we masked any low-SNR ($\rm SNR<3$) pixels in the final data. We used a noise map that includes both the background noise and Poisson noise for each pixel. The background noise $N_{bkg}$ is the standard deviation statistical value of a $12\times12$ pixel square area, and the square of Poisson noise equals the image value: $N^2_{Poisson}=N_{photon}$. The total noise map includes both of Poisson and background noise: $N_{tot} =\sqrt{N^2_{bkg}+N^2_{Poisson}} $. The central and outside regions far away from the arc were masked due to the low SNR.  Thus, we get the lensing arc-only image of the system. The lower left panel of Figure~\ref{fig:subhalo} shows the final data of the strong lensing arc of the JVAS B1938+666 system.

\subsection{Lens Modelling}
\label{sec:subhalo}

After data reduction, our modelling proceeded in two steps to characterize the lens system and search for potential dark matter substructure: Step (1) initial main lens and source model fitting; Step (2) joint fitting including convergence perturbation field.

In Step (1), we employed \href{https://gitlab.mpcdf.mpg.de/ift/lenscharm}{$\emph{LensCharm}$}\citep{lenscharm} to find the main galaxy convergence and source light models by fitting the Keck Kp-band arc image of JVAS B1938+666. The source light model is a parametric S\'ersic model \citep{1963BAAA....6...41S} times a non-parametric perturbation field. The main lens galaxy convergence model is a single smooth pseudo-isothermal ellipsoid (PIE) profile \citep{2008gady.book.....B}. The details of the parameters are in Appendix~\ref{app:lenscharm}.

In Step (2), the best-fitting parameters for $\kappa_{\mathrm{main}}$ and the source models $S_{\mathrm{source}}$ obtained in Step 1 were then used as initial conditions for a subsequent joint modelling phase.  We introduced a non-parametric subhalo convergence perturbation $\Delta\kappa_{\rm sub}$ into the optimized macromodel. The total convergence becomes:
\begin{equation}
    \kappa_{\rm tot}(\boldsymbol{\theta}) = \kappa_{\rm main}(\boldsymbol{\theta}) + \Delta\kappa_{\rm sub}(\boldsymbol{\theta})
\end{equation}
where $\boldsymbol{\theta}$ represents image-plane coordinates. The $\Delta \kappa (\theta)$ 
of JVAS B1938+666 is assumed as a correlated field model (see Section~3.2 of \cite{lenscharm} for more details) in \href{https://gitlab.mpcdf.mpg.de/ift/lenscharm}{$\emph{LensCharm}$}. The correlated field model leverages Gaussian processes while dynamically controlling their power spectrum through an integrated Wiener process \citep{2022PLoSO..1775011G,2022NatAs...6..259A}, enabling adaptive nonparametric reconstructions.  
There are nine free parameters in the two groups used for the correlated field model (see \href{https://ift.pages.mpcdf.de/nifty/user/niftyre_getting_started_4_CorrelatedFields.html}{NIFTY-CorrelationField} or Section~3.2 of \cite{lenscharm}). The prior ranges of the parameters for the correlated field model are listed in Appendix~\ref{app:lenscharm}.
In the fitting process with a convergence perturbation field, the parameters of both the primary galaxy and source models are treated as free parameters.

Using the best-fit main lens convergence and source light model from Step (1) as the initial condition, hence, we performed joint optimization of the main lens convergence, source light, and the non-parametric perturbation field. This joint fitting required multiple iterations to achieve convergence due to the complex degeneracies between these components. 
We calculated the value of $\chi^2$ responding to the output of the model for each iteration to ensure that the value of $\chi^2$ decreases with the iteration process and finally stabilizes near the lowest value, indicating that the fit gets better and better with the iteration and finally reaches the optimal solution. The iteration with the minimum $\chi^2$ value was selected as our final best-fit solution. From this optimal solution, the subhalo's contribution to the convergence ($\Delta\kappa(\theta)$) was robustly inferred. 

Note that in the two-step approach of fitting the main halo and light source first, and then searching for subhalo contributions, We assumed a priori the existence of a subhalo in JVAS B1938+666 based on the conclusions of \cite{2012Natur.481..341V} and \cite{2012MNRAS.424.2800L}.

Finally, we achieved the detection of the subhalo in the strong lensing system JVAS B1938+666. The strong lens modelling results are shown in Figure~\ref{fig:subhalo}. The image axes in Figure~\ref{fig:subhalo} are scaled in arcseconds. The upper panel illustrates the best-fit source surface brightness reconstruction (top left), convergence (top centre), and the added convergence perturbation attributable to the subhalo (top right). The bottom panel displays the reduced Keck/NIRC2 Kp-band data for JVAS B1938+666 (bottom left), the best-fit model post-PSF convolution (bottom centre), and the residuals between the observational data and the best-fit model (bottom right).

The reconstructed subhalo density profile is plotted as black points with error bars in Figure~\ref{fig:result}.
A summary of the detected subhalo's mass distribution in JVAS B1938+666 is presented in Table~\ref{tab:data} in Appendix~\ref{app:lenscharm}. We report the subhalo's convergence-derived surface mass density and enclosed mass within annular bins, computed from the reconstructed $\Delta\kappa (\theta)$. The density and mass errors for each radius bin encompass both statistical and systematic components. The statistical error arises from the fluctuations in the subhalo convergence within each radius bin. The systematic errors were estimated using a simulated strong lensing system with a subhalo of mass $ \rm M_{sub} (r<4.6\, kpc) =  2\times 10^8 \, \rm M_{\odot}$ and a lens galaxy of mass $\rm M_{main} (r<4.6 \, kpc) =  10^{10}\, \rm M_{\odot}$. Given the Einstein ring radius of JVAS B1938+666 is $r=6.4$ kpc, the above mock parameters were selected within that radius. The mock image noise was added with a Gaussian noise field identical to that of the current Keck observation of JVAS B1938+666 (see Appendix~\ref{app:sys} for more details on the estimation of the systematic errors). We calculated the relative difference between the input subhalo density simulation values and the best fit to the simulated lens as the relative systematic error. As demonstrated by the simulation tests in Appendix~\ref{app:sys}, we can not only quantify the systematic errors associated with the reconstructed subhalo, but also demonstrate that our program is capable of distinguishing between flat-core and steep NFW subhaloes.

\subsection{Subhalo Density Profile Fitting}

We fitted the density profile of the reconstructed subhalo of JVAS B1938+666 found in Section~\ref{sec:subhalo} to profiles of the dark matter density profiles of $\psi$DM, SIDM, and NFW, resepectively. Subsequently, we compared the best-fit results across these models.

For the $\psi \rm DM$ model, we adopted the widely recognized profile from \citet{2014PhRvL.113z1302S}, which was derived from $\psi \rm DM$ simulations. In the case of the SIDM model, we employed the parametric profile from \citet{2024PhRvD.110j3044Y,2024JCAP...02..032Y,2025arXiv250214964H}, also obtained through simulation fitting. Additionally, we used the NFW profile \citep{1996ApJ...462..563N} as a reference model for comparison with SIDM and $\psi\rm DM$.

The $\psi$DM density profile across different radii was calculated using Equation~(\ref{eq:rho}), while Equation~(\ref{eq:rho_SIDM}) was applied to determine the SIDM density profile. The NFW profile was computed via Equation~(\ref{eq:rhoNFW}).

The dark matter soliton core of $\psi$DM is larger in low-mass halos (as shown in Equation~(\ref{eq:r_c}) or Figure~\ref{fig:rho} in Appendix~\ref{app:FDM_profile}). Thus, low mass subhalos (i.e., $M_{h} < 10^{9} \, \rm M_{\odot}$) have advantages in constraining $\psi$DM particle mass compared to massive halos because of predicted large core size in the case of $\psi \rm DM$ model. 
Because the flat core radius of a massive elliptical galaxy is significantly smaller than that of a low-mass halo under the $\psi$DM scenario (see Equation~(\ref{eq:r_c}) or Figure~\ref{fig:rho} in Appendix~\ref{app:FDM_profile}), and because the pseudo-isothermal ellipsoid (PIE) profile (Equation~(\ref{eq:kappa_PIE}) in Appendix~\ref{app:B.1.1}) includes a free parameter $r_c$ that can accommodate a flat core, the PIE model serves as a suitable main halo model—regardless of whether the central density profile exhibits a core or a cusp. We also tested the simulation system using a $\psi$DM-inspired flat core in the main halo and found that the subhalo profile remains largely unaffected by the presence of such cores.

To estimate the posterior probabilities of the free parameters, we implemented the Markov Chain Monte Carlo (MCMC) method. The likelihood function is expressed as:

\begin{equation}
\mathcal{L}\left(\rho,r \vert \Theta\right) =
\prod_i \frac{\exp \left\{ -\frac{\left[\rho_{\text{obs},i}\left(r_{i} \right)- \rho_{\text{model},i}(r_i;\Theta)\right]^2}{2(\sigma_{\text{obs},i})^2} \right\}}{\sigma_{\text{obs},i}\,\sqrt{2\pi}},
\label{eq:likeli}
\end{equation}
where $\rho_{\text{obs},i}(r_i)$ denotes the subhalo density at radius $r_i$ as listed in Table~\ref{tab:data}, $\rho_{\text{model},i}(r_i; \Theta)$ represents the density profile model with parameter set $\Theta$, and $\sigma_{\text{obs},i}$ is the associated error of the subhalo density profile.

Table~\ref{tab:prior} summarizes the prior distributions for the parameters of the three models under consideration.

\begin{table}
    \centering
    \begin{tabular}{cccc}
    \hline
    \hline
    parameter & prior range & prior type &  posterior \\
    \hline
    \hline
    \multicolumn{4}{c}{$\psi$DM}\\
    \hline
    $\log{(m_{22})}$  & [-3.0, 4.5] & Uniform  &  $0.12^{+0.09}_{-0.08}$ \\
    $\log{(m_h)}$ & [-3.0, 3.0] & Uniform  &  $1.79^{+0.10}_{-0.11}$   \\
    $r_s/r_c$ & [2.7, 3.5] & Uniform  & $2.86^{+0.23}_{-0.12}$ \\
    \hline
    \multicolumn{4}{c}{SIDM} \\
    \hline
    $\log{(\tau)}$ & [-4.0, -0.7] &  Uniform  &  $-2.15^{+0.90}_{-0.90}$ \\ 
    $\log{(\rho_{s,0})}$  & [2.0, 12.0] &  Uniform & $6.62^{+0.31}_{-0.37}$   \\
    $\log{(r_{s,0})}$ & [-5.0,  6.0] & Uniform  &  $0.67^{+0.27}_{-0.16}$ \\
    \hline
    \multicolumn{4}{c}{NFW} \\
    \hline
    $\log{(\rho_0)}$  & [4.0, 12.0] & Uniform  &   $5.97^{+0.08}_{-0.09}$  \\
    $\log{(r_s)}$ & [-5.0,  6.0] & Uniform 
 &  $0.68^{+0.07}_{-0.06}$ \\
    \hline
    \hline
    \end{tabular}
    \caption{The prior ranges of the parameters for fitting the subhalo density profile. The $\psi\rm DM$ mass particle mass is rescaled with a value of $10^{-22} \, \rm eV$ ($m_{22}\equiv m_{\psi} /(10^{-22}\, \rm eV)$) and halo mass of $\psi$DM at radius of 100 kpc is rescaled with a value of $10^8\, \rm M_{\odot}$ ($m_{h} = M_{h,100\, \rm kpc}/ (10^{8}\, \rm M_{\odot})$).}
\label{tab:prior}
\end{table}

\section{Result}
\label{sec:result}

\begin{figure}[ht]
\centering
 \includegraphics[width=1\columnwidth]{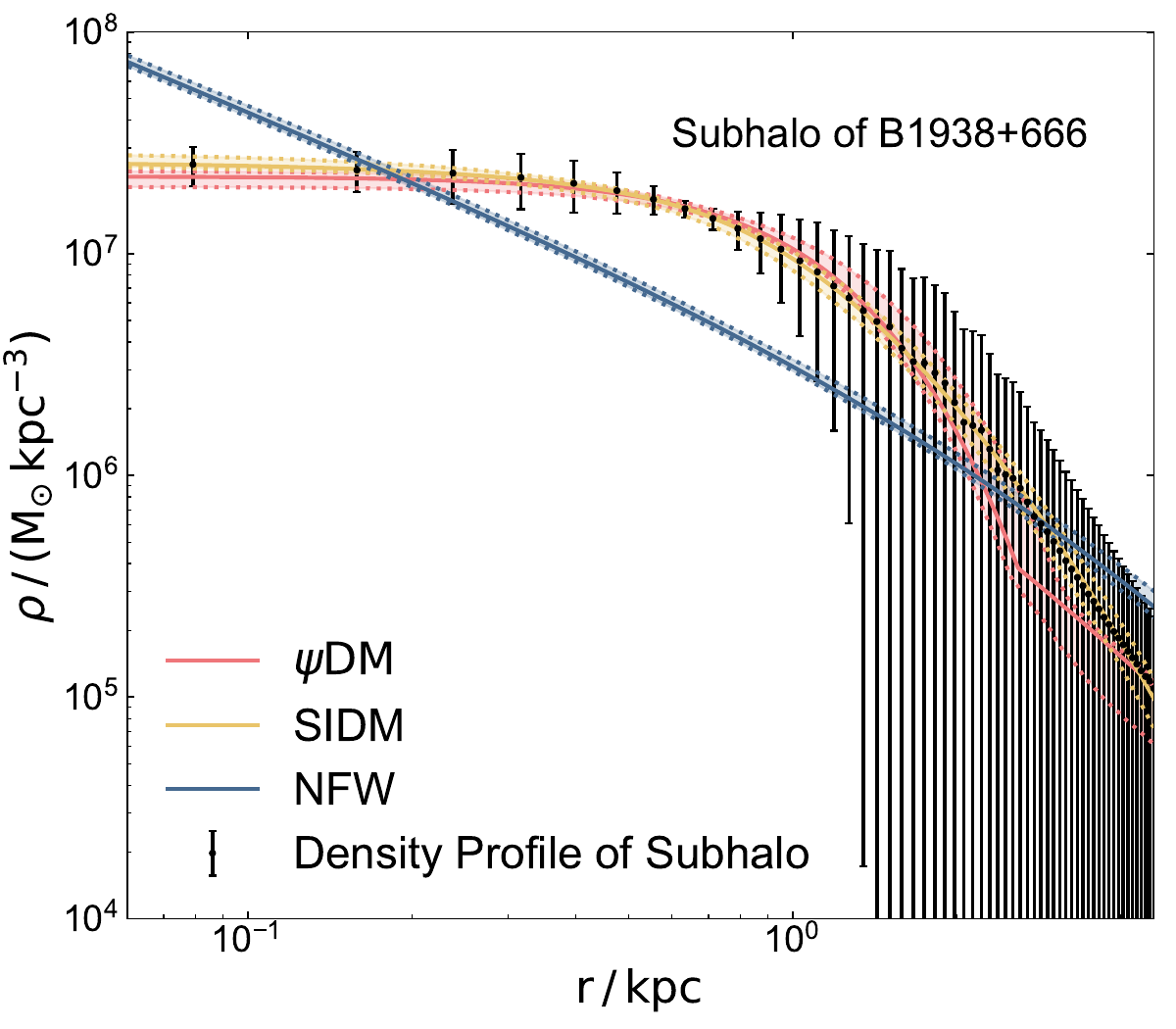}
 \caption{The fitting results of different dark matter density profiles. The data points with error bars are the mass distribution of the subhalo of JVAS B1938+666 listed in Table~\ref{tab:data}. The solid line shows the maximum log-likelihood result of the dark matter density profile model. The dashed band shows the fitted 68\% region of the profile model. The lines in red, yellow, and deep blue are the results of $\psi$DM, SIDM, and NFW models, respectively.  }
 \label{fig:result}
\end{figure}

Figure~\ref{fig:result} presents the results of the dark matter density profile models that were optimised to fit the JVAS B1938+666 sub-halo data. The best-fit models for the subhalo of JVAS B1938+666 are depicted as solid lines, with the dashed regions indicating the 68\% posterior interval. The red, yellow, and deep blue lines correspond to the $\psi$DM, SIDM, and NFW models, respectively. The posterior distributions of the parameters are illustrated in Figure~\ref{fig:posterior} in Appendix~\ref{app:posterior}, and the parameter posteriors are also tabulated in the last column of Table~\ref{tab:prior}.

\begin{table}
    \centering
    \begin{tabular}{cccc}
    \hline
    \hline
     value & $\psi$DM & SIDM & NFW \\
    \hline
    \hline
    $\ln{\mathcal{L}}_{max}$  & -1.39 & -0.05 &  -96.6  \\
    $\Delta \chi^2$ \footnote{The difference of $\chi^2$ from the model with $\psi$DM model: $\Delta \chi^2 = \chi^2_{model} - \chi^2_{\psi\rm DM} $} &  0  &  -2.67  &  +190.4  \\
    $\Delta \rm BIC$ \footnote{The difference of BIC from the model with $\psi$DM. We use it as a proxy of logarithmic Byes factor, $\ln{(BF)} = \Delta {\rm BIC} = {\rm BIC}_{\psi {\rm DM}} - {\rm BIC}_{model} $.  } & 0  & +2.67  &  -186.3 \\
    \hline
    \hline
    \end{tabular}
    \caption{The comparison of the best-fit results of the dark matter density profile models.}
\label{tab:models_likeli}
\end{table}

We also tested the possible stellar mass contribution by adding a Hernquist profile parametrized as in Eq.~(\ref{eq:rho_Hernquist}) to the fitting of the model. We jointly fitted the Hernquist profile in Eq.~(\ref{eq:rho_Hernquist})  and the dark matter profile, with all parameters as free parameters to fit the density profile of the subhalo of JVAS B1938+666. We find no significant baryon contribution ($M_*<10^3 \, \rm M_{\odot}$) in all the fits of the three models. Thus, we show the results without adding the baryon profile. Note that here we only fit the density profile and do not consider the effect of baryon feedback on the dark matter halo profile. The effect of baryon feedback on dark matter halos will be discussed in Section \ref{sec:dis}.

The maximum log-likelihood of the $\psi$DM profile model is $\ln\left({\mathcal{L}_{max,\psi{\rm DM}}}\right)=-1.39$, and that is $\ln\left({\mathcal{L}_{max,\rm SIDM}}\right)=-0.05$ for the SIDM profile. Thus we can calculate the $\Delta$BIC for $\psi$DM as a proxy for the model log evidence. The result of this calculation yields a $\Delta$BIC = 2.67,  corresponding to a Bayes factor $BF = 14.44$. Thus the data moderately favors the SIDM profile model over $\psi\rm DM$. The fitting goodness of the NFW model compared with $\psi$DM is $\Delta \rm BIC = -186.3$, which suggests a significant tension between NFW and data assuming the reconstructed subhalo profile is representative.
The difference between the models can also be found in Table~\ref{tab:models_likeli}.

The current results show that the subhalo of JVAS B1938+666 strong lensing system favours the SIDM or $\psi$DM model. 
The best-fit dark matter particle mass of $\psi$DM is $m_{\psi}=1.3^{+0.3}_{-0.2}\times 10^{-22} \, \rm eV$, which is consistent with the result of anomalies in gravitationally lensed images \citep{2023NatAs...7..736A} or local dwarf galaxies \citep{2023A&A...676A..63B},
but challenged by 
the gravitationally lensed radio jet \citep{2023MNRAS.524L..84P} and the power spectrum of the Lyman-$\alpha$ Forest \citep{2021PhRvL.126g1302R,2017PhRvL.119c1302I}.

\section{Conclusion and Discussion}
\label{sec:dis}

In this work, we reconstructed the subhalo mass distribution of JVAS B1938+666 using Keck observational data and a new method without assuming a specific parametrization for the subhalo mass distribution. We explored the density profile models of dark matter associated with the detected subhalo of JVAS B1938+666, including $\psi$DM, SIDM and NFW. Our analysis shows that the SIDM or $\psi$DM dark matter profile fitted the subhalo density better than NFW with high Bayesian evidences.

In this work, we used a hybrid modelling of \href{https://gitlab.mpcdf.mpg.de/ift/lenscharm}{LensCharm} \citep{lenscharm} to fit the lensing system JVAS B1938+666, including parametric and nonparametric approaches. Although we obtained the profile of the subhalo listed in Table~\ref{tab:data}, it depends on some assumptions, such as the subhalo redshift being consistent with the main lens galaxy, the lens light being removed, and the form of the perturbation field. Different strong gravitational lens modelling programs 
may yield divergent results. The current findings represent only the best-fit solution from \href{https://gitlab.mpcdf.mpg.de/ift/lenscharm}{\emph{LensCharm}}.

\begin{figure}[ht]
\centering
 \includegraphics[width=1\columnwidth]{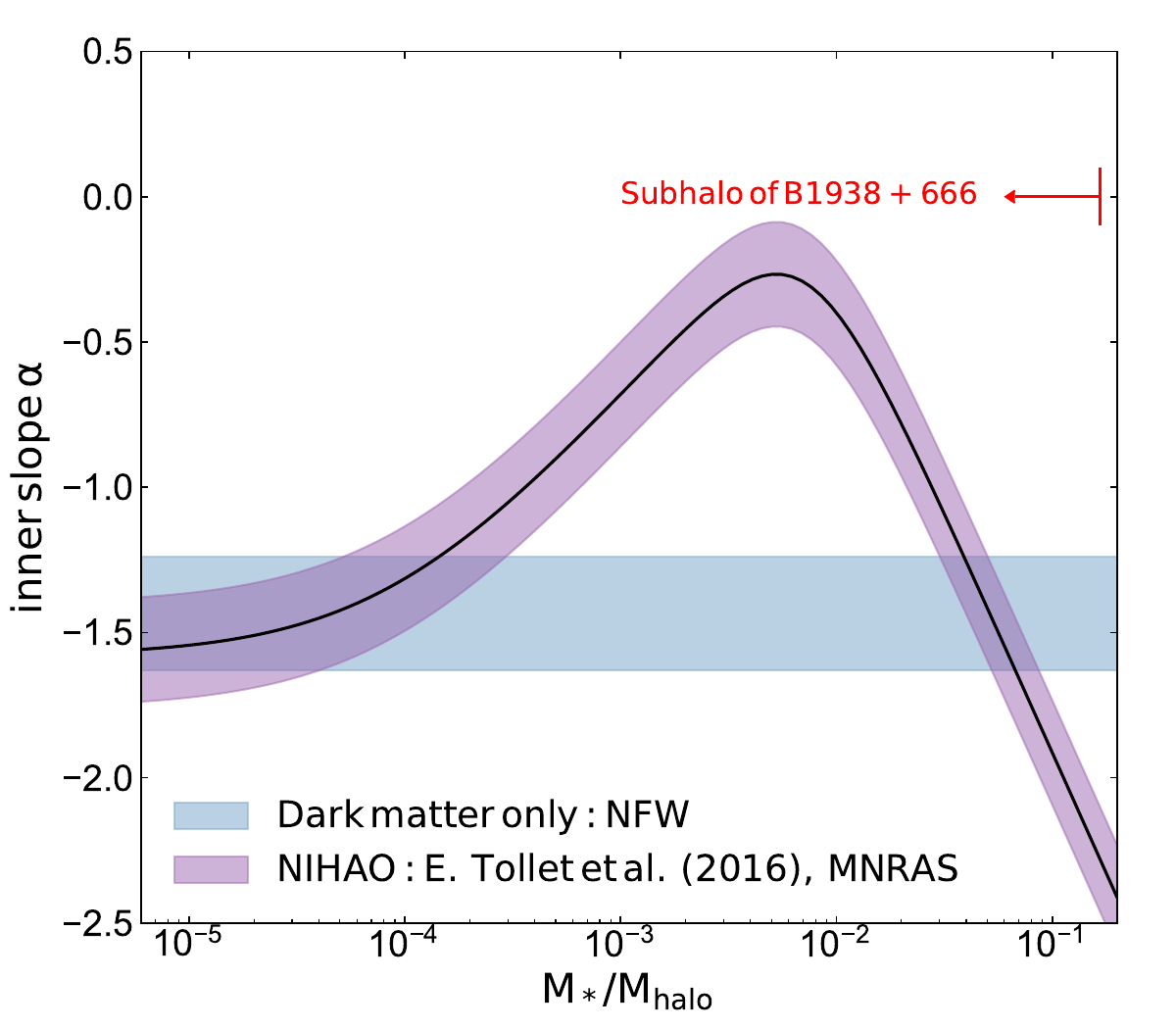}
 \caption{The constraint on the inner slope-ratio of the stellar mass to halo mass relationship of the subhalo of JVAS B1938+666. The red arrow is the upper limit of the subhalo stellar mass-to-halo mass ratio. The purple and blue bands are predictions of inner slopes from the simulations with baryonic feedback \citep{2016MNRAS.456.3542T} and without baryonic feedback \citep{2017ARA&A..55..343B}.   }
 \label{fig:innerSlope}
\end{figure}

The two previous works \citep{2022MNRAS.516..336S,2025A&A...699A.222D} used parameterized subhalo models to obtain steep profile slopes, where \cite{2022MNRAS.516..336S} found it's power-law slope was $\gamma_{pl}=1.96$ and \cite{2025A&A...699A.222D} fitted the power-law slope was $\gamma_{pl}=2.42$.
However, considering models that can give the center a flat profile will lead to different conclusions. For example, \cite{2014MNRAS.442.3598V} uses the extended NFW model, and the center slope parameter can cover the flat and sharp situations. The results show that the center slope of JVAS B1938+666 is not very steep, and the constraint is $\gamma<1.6$. This is consistent with our result. This may indicate that the power law profile is too simplified for the real subhalo, making it difficult to fully fit the overall profile of the subhalo.

Although the dense core of the subhalo is well-fitted by both SIDM and $\psi$DM profiles, baryonic feedback remains a viable alternative core formation mechanism. Driven primarily by supernova explosions, this process generates repeated cycles of gas outflow and inflow, causing fluctuations in the gravitational potential \citep{2010Natur.463..203G,2012MNRAS.421.3464P}. These time-varying potentials transfer energy to dark matter particles through gravitational interactions, redistributing them outward and transforming central cusps into flatter cores \citep{2014Natur.506..171P,2016MNRAS.456.3542T,2017ARA&A..55..343B}. Following the subhalo's discovery in JVAS B1938+666, \citet{2012Natur.481..341V} found no significant associated stellar luminosity, providing only an upper limit. Using this luminosity constraint and assuming $M/L=1$, we derive a corresponding upper limit for $\rm M_*/M_{halo}$, indicated by the red arrow in Figure~\ref{fig:innerSlope}. As shown in Figure~\ref{fig:innerSlope}, this upper limit permits core formation via baryonic feedback, as the NIHAO simulations show core flattening ($\alpha \approx 0$) occurring at $\rm M_*/M_{halo} \sim 0.005$ (purple band). However, baryonic adiabatic contraction could act in opposition to baryonic feedback, potentially suppressing the formation of a flat core \citep{2004ApJ...616...16G,2014MNRAS.441.2986D,2015MNRAS.454.2981C,2020MNRAS.497.2393L}. If its influence is dominant, the observed flat profile may not be fully explained by feedback processes alone, and might instead require explanations of SIDM or $\psi$DM.

Future high-sensitivity instruments like next-generation VLBI arrays or the Extremely Large Telescope (ELT) could detect baryonic components through emission lines from gas or stars within such substructures. Such detections would provide stronger constraints on both dark matter properties and baryonic feedback processes.

Distinguishing subhalos from interlopers (line-of-sight halos) in strong lensing systems at higher redshift planes poses a significant challenge, as noted in recent studies \citep{2024arXiv241108565E}. For the system JVAS B1938+666, analyses by \cite{2022MNRAS.515.4391S,2022MNRAS.516..336S} of Hubble Space Telescope observations suggest that the perturbing mass is a line-of-sight halo at 
$z \sim 1.2$, utilizing the multiplane thin-lens approximation. If confirmed, this classification implies a higher concentration for the perturber compared to a subhalo scenario. 

Specifically, if the line-of-sight halo in JVAS B1938+666 potentially resides at a higher redshift rather than being a subhalo at $z\approx0.881$, the dark matter halo might exhibit a slightly larger mass, though the precise difference in mass remains uncertain. However, its observed flat core makes the structure more readily explained by alternative dark matter models such as SIDM and $\psi$DM. In contrast, the CDM paradigm would require strong stellar feedback mechanisms to account for such a core. Existing observational upper limits on stellar populations, however, indicate that the number of stars might be insufficient to support this explanation, creating a potential tension with the CDM model.
Consequently, these measurements could provide support for SIDM and $\psi$DM models, which naturally predict the formation of flat cores in dark matter structures. The conclusion that SIDM or $\psi$DM might better explain the data compared to the traditional NFW profiles used in CDM remains plausible but requires further confirmation.

\cite{2024MNRAS.52710480N} searched dark matter subhaloes in Hubble Space Telescope imaging of 54 strong lenses. However, the subhaloes detected with high signal-to-noise ratios (SNR) by \cite{2024MNRAS.52710480N} have masses exceeding $>10^9\, \rm M_{\odot}$. Upcoming observations from advanced instruments such as the James Webb Space Telescope (JWST), 
the Extremely Large Telescope (ELT), the Large Synoptic Survey Telescope (LSST), DESI Legacy Survey Imaging, and the China Space Station Telescope (CSST) are eagerly anticipated to furnish more nuanced insights into the distribution of dark matter within subhalos. Furthermore, innovative methodologies, including machine learning techniques \citep[e.g.,][]{2020PhRvD.101b3515D,2023A&A...672A.123H}, could be harnessed to directly analyze large samples of dark matter substructure in strong lensing images. The application of these extensive datasets holds promise for significantly bolstering the constraints on the properties of dark matter particles.



\section{Acknowledgments}
We thank Matteo Guardiani, Julian R{\"u}stig, Torsten En$\ss$lin, Jiang Chang, Yuan-Lin Gong, Yong-Jia Huang, Zhi-Ping Jin, Ke-Yu Lu, Guan-Wen Yuan and Hao Zhou for the helpful discussion. This research has made use of the Keck Observatory Archive (KOA), which is operated by the W. M. Keck Observatory and the NASA Exoplanet Science Institute (NExScI), under contract with the National Aeronautics and Space Administration. This work is supported in part by the National Key R\&D Program of China (2022YFF0503304) and by the NSFC (No. 12233011).

%

\vspace{5mm}


\software{\href{https://gitlab.mpcdf.mpg.de/ift/lenscharm}{LensCharm} \citep{lenscharm}, astropy \citep{2013A&A...558A..33A,2018AJ....156..123A}, emcee \citep{2013PASP..125..306F}, matplotlib \citep{Hunter:2007}, numpy \citep{harris2020array}, scipy \citep{2020SciPy-NMeth}.}



\appendix

\section{Density Profile Models of Dark Matter}
\label{app:DM_profile}

\subsection{$\psi$DM Density Profile Model}
\label{app:FDM_profile}

The soliton core formation in $\psi$DM is due to the uncertainty principle creating quantum pressure that counteracts gravity, resulting in a flat central density profile \citep{2000ApJ...534L.127P,2000PhRvL..85.1158H,2014MNRAS.437.2652M,2000PhRvL..84.3760S}.
The soliton density profile as a function of the dark matter mass $m_{\psi}$ is found to be \citep{2014PhRvL.113z1302S}
\begin{equation}
 \rho_c(x)=\frac{1.9\times 10^7 \left(\frac{m_\psi}{10^{-22} \mathrm{eV}}\right)^{-2}\left(\frac{x_c}{\mathrm{kpc}}\right)^{-4}}{a \left[1+9.1 \times 10^{-2}\left(\frac{x}{x_c}\right)^2\right]^8} M_{\odot} \, \mathrm{kpc}^{-3} .
 \label{eq:rhoc}
\end{equation}

The characteristic scale of the soliton core $x_c$ in Equation~(\ref{eq:r_c}) is related to the cosmological scale factor $a$ and the radius of the soliton core $r_c$ as \citep{2014PhRvL.113z1302S}
\begin{equation}
 x_c = r_c/a,
\end{equation}

\begin{equation}
 r_c = \frac{1.6 a ^{1/2}}{ \left(\frac{m_{\psi}}{10^{-22}\, \rm eV}\right) \left(\frac{\zeta(z)}{\zeta(0)}\right)^{1/6} \left( \frac{M_h}{10^9\, \rm M_{\odot}} \right)^{1/3} } \, \rm kpc,
 \label{eq:r_c}
\end{equation}

\begin{equation}
 \zeta(z) = \frac{18\pi ^2 + 82 \left(\Omega_m(z)-1\right) - 39 \left(\Omega_m(z)-1\right)^2}{\Omega_m(z)}.
\end{equation}
Here, $\Omega_m(z)$ is the matter fraction in the total energy density at redshift $z$, and $M_h$ is the total halo mass of the observed dark matter halo. 

The density of the outer region of the $\psi$DM halo soliton is consistent with the Navarro-Frenk-White (NFW) profile \citep{1996ApJ...462..563N}
\begin{equation}
\rho_{\rm NFW}(r) = \frac{\rho_0}{ \frac{r}{r_s}\left( 1+\frac{r}{r_s} \right)^2 },
\label{eq:rhoNFW}
\end{equation}
where $\rho_0$ is the dark matter density at the scale $r_s$ estimated from the soliton core profile
$\rho_{0} = 4 \rho_c (r_s)$. The scale is fixed as $r_s/r_c = 3.3$ according to simulation \citep{2021PhRvD.103j3019C,2017MNRAS.471.4559M}.

The NFW outer profile of $\psi$DM is not only found in simulation by \cite{2014NatPh..10..496S,2014PhRvL.113z1302S},
but also favored in different halo fits.  
\cite{2023A&A...676A..63B} found that the soliton+NFW model offers an excellent fit to the rotation curves of the dwarf galaxy sample, with inferred axion masses clustering around a relatively narrow range of values $m_{\psi} \sim (1 - 5) \times 10^{-23}$ eV. Similar conclusions are also drawn by \cite{2023arXiv230200181P} in the dwarf galaxy sample.


The full dark matter density profile of a dark matter halo can be expressed as 
\begin{equation}
 \rho(r)=
\left\{ 
    \begin{array}{lc}
        \rho_{c}(r), & r \leqslant r_s, \\
        \rho_{\rm NFW}(r), & r>r_s, \\
    \end{array}
\right.
\label{eq:rho}
\end{equation}   
where $\rho_{c}(r)$ and $\rho_{\rm NFW}(r)$ represent the soliton core density profile in Equation~(\ref{eq:rhoc}) and the NFW outskirt density profile in Equation~(\ref{eq:rhoNFW}), respectively.

Therefore, the cumulative dark matter mass profile of the halo can be calculated as
\begin{equation}
M(r) = \int^{r}_{0} 4 \pi r^2 \rho(r)\,  dr.
\label{eq:M_r}
\end{equation}

The cumulative mass in Equation~(\ref{eq:M_r}) can be straightforwardly calculated using numerical methods once the analytic Equations~(\ref{eq:rhoc}-\ref{eq:rho}) are considered. We calibrate the cumulative halo mass by assuming $M(<\rm 100 \, kpc)$ is equal to the total halo mass.

\begin{figure*}
\centering
 \includegraphics[width=0.49\columnwidth]{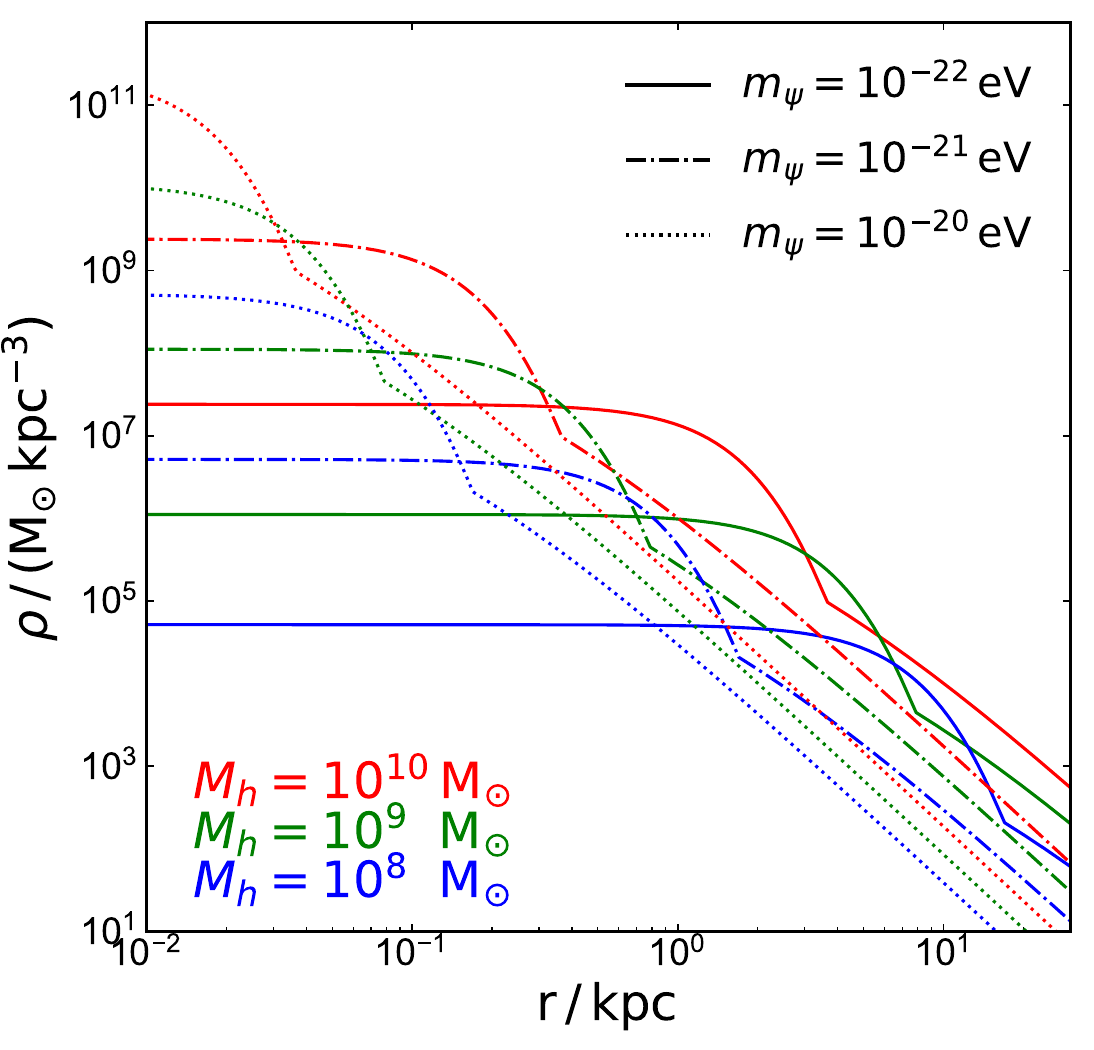}
 \includegraphics[width=0.49\columnwidth]{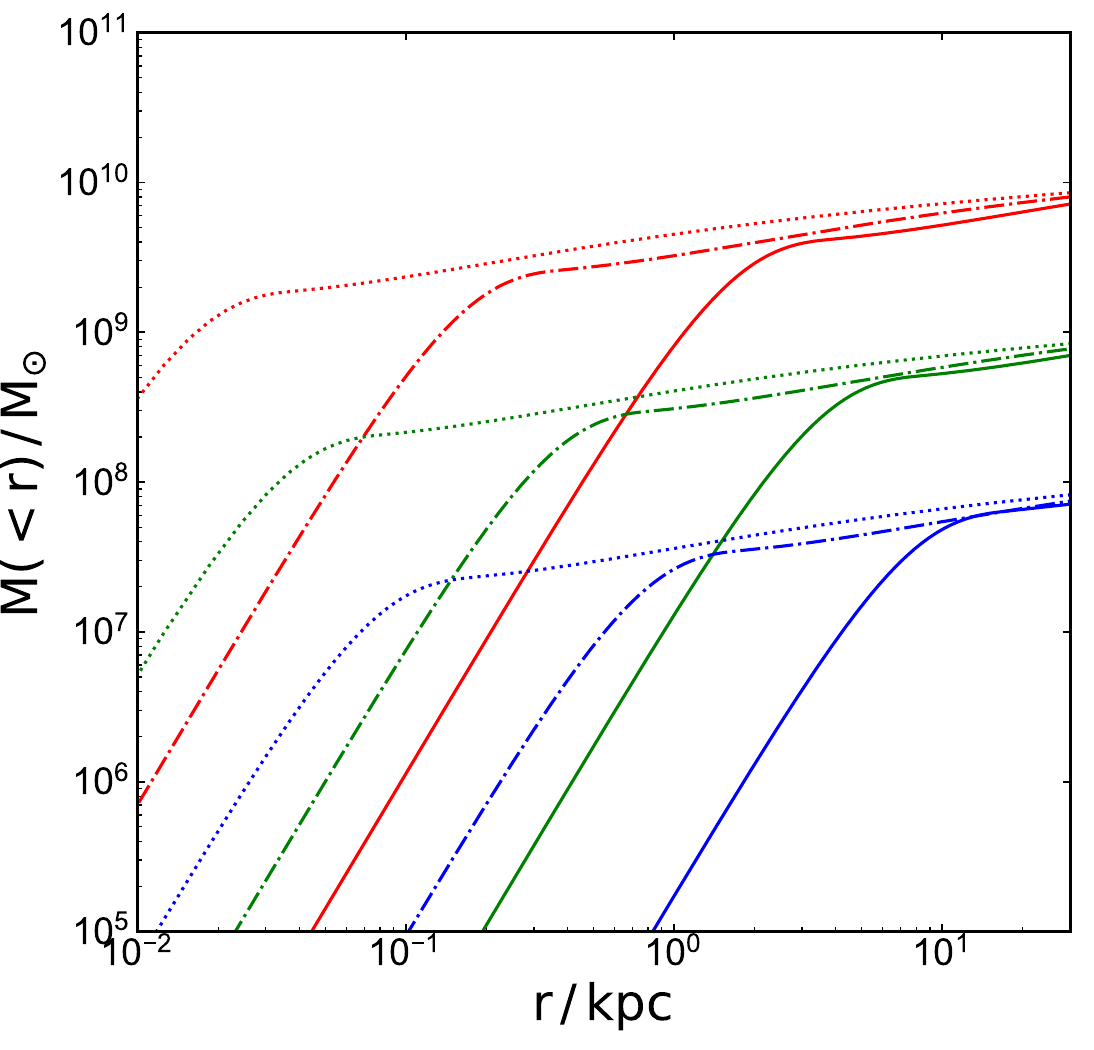}
 \caption{Left panel: Dark matter halo density profile with different dark matter masses ($10^{-22}, 10^{-21}, 10^{-20}$ eV) and halo masses ($10^{8}, 10^{9}, 10^{10}$ $M_{\odot}$) at redshift $z= 0.881$. 
 Right panel: Cumulative dark matter mass profile of haloes with different dark matter masses ($10^{-22}, 10^{-21}, 10^{-20}$ eV) and halo masses ($10^{8}, 10^{9}, 10^{10}$ $M_{\odot}$) at redshift $z= 0.881$.
 The starting radius scale of the outer NFW profile is fixed as $r_s = 3.3 r_c$ in these plots. The density model combines soliton and NFW profiles in the inner and outer regions, respectively.}
 \label{fig:rho}
\end{figure*}

Figure~\ref{fig:rho} shows dark matter density and cumulative mass profiles with varying parameters. The left panel of Figure~\ref{fig:rho} displays the dark matter density profiles. 

The soliton core appears as the smooth core shape in the inner region smaller than the turning point on the plots. The larger region that exceeds the turning point is the NFW profile because the uncertainty principle is not significant when the scale is much larger than the soliton core size. The right panel of Figure~\ref{fig:rho} shows that the dark matter mass distribution of a halo is sensitive to dark matter particle mass in the theory of $\psi \rm DM$. 

Furthermore, the dark matter soliton core is more detectable in the low-mass halos when the dark matter particle mass is fixed. Dwarf galaxies with low masses (i.e., $M_{h} < 10^{9} \, \rm M_{\odot}$) have advantages in constraining $\psi$DM particle mass compared to massive halos because of their large core size in the $\psi \rm DM$ model. 

The same conclusion can be readily inferred from Equation~(\ref{eq:r_c}). For instance, the dark matter core size $r_c$ is larger in lighter mass dark matter halos when the dark matter particle mass is fixed.

\subsection{SIDM Profile Model}
\label{app:SIDM_profile}

The analytic density profile of SIDM has been established in the previous works of
\cite{2024PhRvD.110j3044Y,2024JCAP...02..032Y,2025PDU....4701807Y,2025arXiv250214964H}. The subhalo density profile is described as \citep{2024JCAP...02..032Y}:

\begin{equation}
 \rho(r)= \frac{\rho_s}{\frac{\left( r^{\beta}+r_c^{\beta} \right)^{\frac{1}{\beta}}}{r_s} \left( 1+\frac{r}{r_s}\right)^2 \left[ 1+ \left( \frac{r}{r_t} \right)^{2-u} \right]^{1+3u}}.
\label{eq:rho_SIDM}
\end{equation}   

Here $\rho_s$ and $r_s$ are respectively scale density and radius, $r_c$ is the core radius, $\beta$ is the parameter for controlling the transition of the inner core and outer NFW profile, $u$ is the numerical factor controlling the logarithmic slope of the truncated density profile in the outer region, $r_t = 1.53$ kpc is the tidal radius of JVAS B1938+666. 

According to \cite{2024JCAP...02..032Y}, the evolution of the subhalo density profile parameters $\rho_s$, $r_s$ and $r_c$ can be written as the function related to rescaled evolution time $\tau$
\begin{equation}
\tau \equiv \frac{t}{t_c},
\label{tau}
\end{equation}
where $t$ is the age of the dark matter halo, and $t_c$ is the time of core collapse of the halo

\begin{equation}
t_c = \frac{150}{C} \frac{1}{\left( \sigma_{\text{eff}}/m\right) \rho_{s,0}\, r_{s,0}} \frac{1}{\sqrt{4\pi G \rho_{s,0}}} ,
\label{tc}
\end{equation}
where $C$ is a constant that can be fixed as 0.75 (or 0.9) \citep{2024JCAP...02..032Y,2024PhRvD.110j3044Y}, $\frac{\sigma_{\text{eff}}}{m}$ is the effective cross-section of dark matter interactions.  We also take the emperical relations, including the core density and radius \citep{2024JCAP...02..032Y}:
\begin{equation}
\rho_s = \rho_{s,0}\left( 2.033+0.7381\tau + 7.264 \tau^5 - 12.73 \tau^7 + 9.915 \tau^9 + \left( 1-2.033 \right)\left(  \ln{0.01} \right)^{-1} \ln{ \left( \tau +0.001 \right)}  \right),
\label{rho_s}
\end{equation}

\begin{equation}
r_s = r_{s,0}\left(  0.7178 - 0.1026\tau + 0.2474 \tau^2 - 0.4079 \tau^3 + \left( 1-0.7178 \right)\left( \ln{0.001} \right)^{-1} \ln{\left( \tau +0.001\right)} \right),
\label{r_s}
\end{equation}

\begin{equation}
r_c = r_{s,0}\left( 2.555\sqrt{\tau} - 3.632 \tau + 2.131 \tau^2 - 1.415 \tau^3 + 0.4683 \tau^4  \right).
\label{r_c}
\end{equation}

Thus, there are three free parameters in the SIDM profile model: $\rho_{s,0}$, $r_{s,0}$ and $\tau$.

Due to the self-interaction of dark matter particles, the dark matter halo will have a significant evolutionary status at different times. In particular, in the simulation, we can see the process of the dark matter halo forming a core and then collapsing to a cuspy one \citep{2000PhRvL..84.3760S,2000ApJ...543..514K,2018PhR...730....1T,2022MNRAS.517.3045C,2025PhRvD.111f3001Z}. The different evolutionary stages of these low-mass dark matter halos form the diversity of dark matter halo profiles, which can be used to explain the previous observations of dwarf galaxies \citep{2000ApJ...543..514K,2017PhRvL.119k1102K,2018PhR...730....1T,2019PhRvD.100f3007Z,2019PhRvX...9c1020R,2025arXiv250402303Y}. The above Equation~(\ref{eq:rho_SIDM}) from \cite{2024JCAP...02..032Y} is related to the evolutionary time $\tau$.

\subsection{Baryon distribution profile}

The baryon distribution can be described as a Hernquist profile \citep{2014PhRvL.113b1302K,2023MNRAS.521.4630J}:

\begin{equation}
 \rho(r)= \frac{M_b/2\pi r_0^3}{ \frac{r}{r_0}\left( 1+ \frac{r}{r_0} \right)^{3}}.
\label{eq:rho_Hernquist}
\end{equation}   

Here, $M_b$ is the total baryon mass in the halo, and $r_0$ is the transition radius of the baryon model.

\section{ Dark matter distribution of Subhalo in JVAS B1938+666}
\label{app:lenscharm}

\subsection{Subhalo Reconstruction with LensCharm}

Here, we summarize the parameters in the strong lens modeling process. The parameters of the main lens model and the source light model are listed  in Appendix:\ref{app:B.1.1} and Appendix:\ref{app:B.1.2}, resepectively. We introduce the model parameters of the subhalo and the confidence level of the detected subhalo in Appendix:\ref{app:B.1.3}.

In this work, we found the lens light model of the central galaxy by fitting two free S\'ersic profiles \citep{1963BAAA....6...41S} to the stacked Keck image (masking lensed arcs), then subtracting the best-fit model from the stacked Keck image to produce the final image in Figure \ref{fig:subhalo}. After subtracting the lens light, we masked any low-SNR ($\rm SNR<3$) pixels in the final data. As shown in Figure~\ref{fig:subhalo} in Section~\ref{sec:data}, the central and outside regions far away from the arc were masked due to the low SNR.  Thus, we get the lensing arc-only image of the system. 

So, we excluded potential pollution to subhalo reconstruction from the modelling of the lens light. We use the Python open code \href{https://gitlab.mpcdf.mpg.de/ift/lenscharm}{\emph{LensCharm}}\citep{lenscharm} model the parametric main lens galaxy convergence, source light and non-parametric subhalo convergence.

We model the main lens+subhalo with two steps:

(1) We first fit a mass model that only includes the main lens galaxy mass, and a source light model includes the parametric S\'ersic model times the non-parametric perturbation field. 

(2) Then, the results of the main lens + source light models are used as initial parameters, and we add a non-parametric perturbation field to capture the potential contributions of subhalo mass.

\subsubsection{Main lens galaxy convergence model}
\label{app:B.1.1}

We use a single smooth pseudo-isothermal ellipsoid (PIE) profile \citep{2008gady.book.....B} to fit the main lens galaxy convergence in the system. Because the main lens galaxy is a massive elliptical galaxy, so the main lens galaxy model will not show a large core when the dark matter model is $\psi$DM (see Equation~(\ref{eq:r_c}) or Figure~\ref{fig:rho}). So the PIE model can be used to fit main lens even if the dark matter is $\psi$DM.

The PIE model in \href{https://gitlab.mpcdf.mpg.de/ift/lenscharm}{\emph{LensCharm}}\citep{lenscharm} code is described as:
\begin{equation}
    \kappa(r)=\frac{b}{2}\left(r^2+r_{\mathrm{c}}^2\right)^{-\frac{1}{2}},
    \label{eq:kappa_PIE}
\end{equation}

\begin{equation}
    r=\left\|\left(\begin{array}{cc}
q \cos \theta & -q \sin \theta \\
\sin \theta & \cos \theta
\end{array}\right)\binom{x-x_c}{y-y_c}\right\|_2,
\label{eq:radius_elliptical}
\end{equation}
where $r$ is an elliptical radius defined by Equation~(\ref{eq:radius_elliptical}), $r_c$ is the
core radius, $b$ is the controlling factor of the dimensionless convergence of the main lens galaxy, $q$ is a parameter that steers the ellipticity of the profile, $\theta$ is the inclination angle of the main lens galaxy, $x_c$ is the centre position of the main lens galaxy in the x-axes and $y_c$ is the centre position of the main lens galaxy in the y-axes.  

The $\kappa$ in Equation~(\ref{eq:kappa_PIE}) is a dimensionless value, which is from the surface density 
$\Sigma$ normalized by the lens critical density of dark matter $\Sigma_{\text {crit }}$:
\begin{equation}
    \kappa(r)=\frac{\Sigma(r)}{\Sigma_{\text {crit }}},
    \label{eq:kappa_to_Sigma}
\end{equation}
\begin{equation}
    \Sigma_{\text {crit }}=\frac{c^2}{4 \pi G} \frac{D_s}{D_{ls} D_l},
    \label{eq:Sigma}
\end{equation}
where $c$ is the speed of light, $G$ is the gravitational constant, $D_s$ is the angular diameter distance from the observer to the source plane, $D_{l}$ is the angular diameter distance from the observer to the lens plane, $D_{ls}$ is angular diameter distance from the lens plane to the source plane.

The parameters prior of the PIE main lens galaxy model of JVAS B1938+666 in this work are listed as follows:

\begin{itemize}
    \item $b$: A lognormal distribution with a mean value of 0.5 and $1\sigma$ width of 0.05.
    \item $r_s$: A uniform distribution from $10^{-12}$ to $4\times 10^{-2}$ arcseconds.
    \item $x_c$: A normal distribution with a mean value of $-3\times10^{-2}$ arcseconds and $1\sigma$ width of $5\times10^{-3}$ arcseconds.
    \item $y_c$: A normal distribution with a mean value of $-4\times10^{-2}$ arcseconds and $1\sigma$ width of $5\times 10^{-3}$ arcseconds.
    \item $q$: A uniform distribution from 0.8 to 1.2.
    \item $\theta$: A normal distribution with a mean value of 0.58 and $1\sigma$ width of 1.
\end{itemize}

Additionally, we used a shear model to correct some effects of the galaxy's envoirement. The shear model parameters are:

\begin{itemize}
    \item strength: A normal distribution with a mean value of 0.035 and $1\sigma$ width of 0.02.
    \item $x_c$: A normal distribution with a mean value of $-3\times10^{-2}$ arcseconds and $1\sigma$ width of $5\times10^{-3}$ arcseconds.
    \item $y_c$: A normal distribution with a mean value of $-4\times10^{-2}$ arcseconds and $1\sigma$ width of $5\times 10^{-3}$ arcseconds.
    \item $\theta$: A normal distribution with a mean value of 0.0 and $1\sigma$ width of 10.
\end{itemize}

\subsubsection{Source light model}
\label{app:B.1.2}

We use hybrid models including a parametric model and a nonparametric field model to fit the source light in \href{https://gitlab.mpcdf.mpg.de/ift/lenscharm}{\emph{LensCharm}}\citep{lenscharm}.

The source light model of JVAS B1938+666 is assumed as a Mat\'ern-kernel correlated field (see Section~3.2.2 of \cite{lenscharm} for more details) times a S\'ersic model \citep{1963BAAA....6...41S}. The optimise iterations of the non-parametric correlated field are designed by the nonparametric models for Gaussian
processes developed by \citealt{2022PLoSO..1775011G,2022NatAs...6..259A}. 

There are nine free parameters in the two groups used for the Mat\'ern-kernel correlated field source light model (see \href{https://ift.pages.mpcdf.de/nifty/user/niftyre_getting_started_4_CorrelatedFields.html}{NIFTY-CorrelationField} or Section~3.2.2 of \cite{lenscharm} for more details). We set the parameters prior as follows: 
\begin{itemize}
    \item amplitude: The \emph{offset\_mean} = 1.5 is used to control the mean amplitude offset of the field.  The \emph{offset\_std} = [0.1, 0.001] is used to control the amplitude variations of the offset of the field. The two values are the upper and lower bounds of the offset variations.
    \item fluctuations: The \emph{scale}=[0.4, 0.4] is used to control the fluctuation scale of the Mat\'ern-kernel process.  The \emph{cutoff}=[0.1, 0.01] is used to control the cutoff wavelength of the spatial correlation power spectrum. The \emph{loglogslope}=[-2.5, $10^{-2}$] is used to control the spectral index of the corresponding power spectrum.
\end{itemize}

The S\'ersic model \citep{1963BAAA....6...41S} of the source light is described as:
\begin{equation}
    I(r)=I_e \exp \left(-b_n\left(\left(\frac{r}{r_e}\right)^{1 / n}-1\right)\right).
\end{equation}
There are six parameters for S\'ersic model \citep{1963BAAA....6...41S} of the source light of JVAS B1938+666. We set the prior of the parameters as follows:
\begin{itemize}
    \item $I_e$: A lognormal distribution with a mean value of 0.03 $\rm e^- /s$ and $1\sigma$ width of 0.01 $\rm e^- /s$.
    \item $r_e$: A uniform distribution from $0.05$ to $0.15$ arcseconds.
    \item $n$: A uniform distribution of the S\'ersic index from 1.1 to 3.1.
    \item $x_c$: A normal distribution with a mean value of $-10^{-2}$ arcseconds and $1\sigma$ width of $5\times 10^{-3}$ arcseconds.
    \item $y_c$: A normal distribution with a mean value of $10^{-2}$ arcseconds and $1\sigma$ width of $5\times 10^{-3}$ arcseconds.
    \item $q$: A uniform distribution from 0.8 to 1.2.
    \item $\theta$: A normal distribution with a mean value of 0 and $1\sigma$ width of 1.
\end{itemize}

\subsubsection{Subhalo Model and Detection Evidence}
\label{app:B.1.3}

For non-parametric reconstruction of the subhalo, we use the correlated field model to fit the additional convergence field after iterations of the main lens galaxy model. It should be noted that after adding the perturbation field, the parameters of the original main lens galaxy and source light are also fitted as free parameters, and the parameters without the perturbation field are used as initial condition inputs here.

There are nine parameters in the two groups that are employed in the correlated field model:
\begin{itemize}
    \item amplitude: The \emph{offset\_mean} = -8 is used to control the mean logarithmic amplitude of the field.  The \emph{offset\_std} = [$10^{-8}$, $10^{-16}$] is used to control the amplitude offset and the variations of the field. 
    \item fluctuations: The \emph{fluctuations}=[1.0, $10^{-2}$] is used to control the fluctuations of the correlation field and the variations of the power spectra.  The \emph{loglogavgslope}=[-3.0, 0.5] is used to control the logarithmic slopes of the power spectra of the fluctuations and the variations of these slopes. The \emph{flexibility}=[0.5, 1.0] is used to determine how strong the power spectrum varies besides the power law.
\end{itemize}

To estimate the significance of the dark matter subhalo, we use the Bayes information criterion \citep{1978AnSta...6..461S}:
\begin{equation}
{\rm{BIC}}=-2 \ln \left(\mathcal{L}_{\text{lenscharm}}\right) + k\, \ln \left(n \right).
\label{eq:BIC}
\end{equation}   

The use of BIC can provide a comparison between different models, even if they have different numbers of free parameters. Here $k=9$ is the number of free parameters of the model with subhalo convergence perturbation. The subhalo convergence perturbation is modelled with the fluctuation power spectrum of an information field theory \citep{2009PhRvD..80j5005E} that was adopted by \href{https://gitlab.mpcdf.mpg.de/ift/lenscharm}{\emph{LensCharm}}. Here, $n=150\times 150$ is the total number of data pixels used in the fitting. 
The evidence of the subhalo is 
\begin{equation}
\Delta{\rm BIC} = {\rm BIC}_{main+subhalo} - {\rm BIC}_{main\, galaxy} = -129.10,
\label{eq:deltaBIC}
\end{equation} 
equating to a signal detection confidence of approximately \(10\, \sigma\) if the posterior probability
distribution function was the Gaussian distribution. Even though we know the posterior is likely non-Gaussian, we can use the BIC as a proxy for model comparison. Under this assumption,  the significance of subhalo is similar to the $\sim12 \sigma$ previously analyzed by \cite{2012Natur.481..341V,2025A&A...699A.222D}.

\subsection{Systematic Error Estimation of Subhalo Reconstruction with LensCharm} \label{app:sys}

In order to estimate the possible system errors and test whether the sub-halo systems of different dark matter models with similar signal-to-noise ratios can be successfully distinguished in our method, we simulated the two dark matter sub-halos of NFW and FDM respectively, and we considered the same noise as the observation level to generate the mock images corresponding to different dark matter models, and fitted the simulated images. Here we consider two sets of subhalos simulating NFW and $\psi$DM. 
\begin{figure*}[htbp]
\centering
 \includegraphics[width=1\columnwidth]{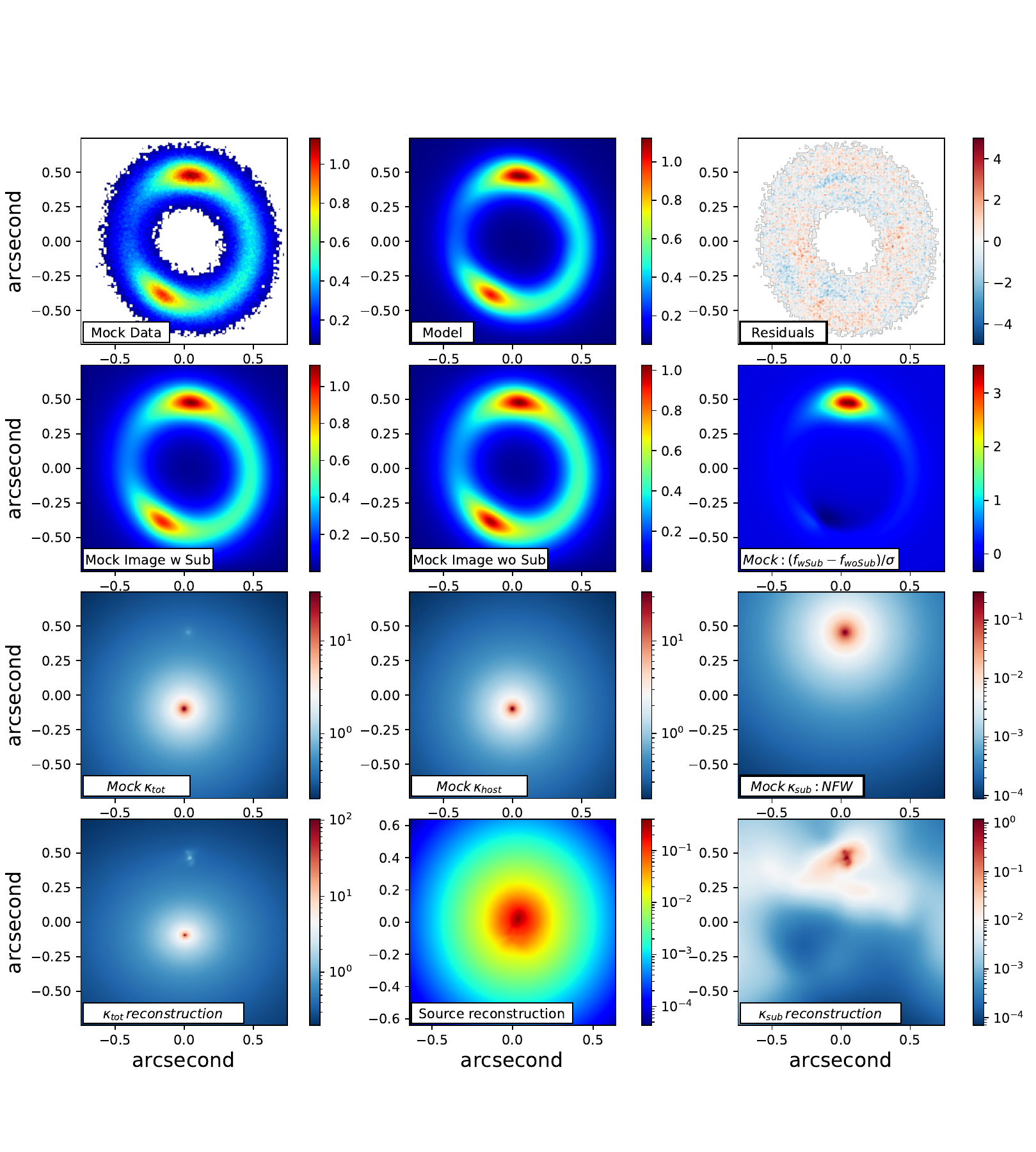}
 \caption{ The mock system with a NFW subhalo. The mock system includes a NFW subhalo with a similar mass and signal-to-noise ratio with the JVAS B1938+666 Keck data. The x and y axes of the images are calibrated in arcseconds. The strong lensing model was constructed using the Python-based tool \href{https://gitlab.mpcdf.mpg.de/ift/lenscharm}{\emph{LensCharm}} \citep{lenscharm}. }
 \label{fig:subhalo_mock_syserr}
\end{figure*}

As shown in Figure \ref{fig:subhalo_mock_syserr}, we have added a NFW subhalo to a simulated strong gravitational lens system similar to JVAS B1938+666. The redshifts of the source and lens of the mock system are also fixed as same as the JVAS B1938+666. The simulated NFW subhalo is set in the position of the top of the north arc. To sample the mock test, we set the ellipticities of the simulated main galaxy and subhalo to zero. The enclosed mass of the subhalo is $M(r\leq4 \,\rm{kpc})=3\times 10^8\, M_{\odot}$.

We fitted the mock data with \emph{LensCharm}, the results is shown in Figure~\ref{fig:subhalo_mock_syserr}. The top panel of Figure~\ref{fig:subhalo_mock_syserr} displays the mock data (top left), the reconstructed model (top centre), the residual (top right). The second panel shows the mock image with NFW subhalo (second panel left), the mock image without subhalo (second panel centre), the SNR of difference between mock images with and without subhalo (second panel right). The third panel displays the total mock lens convergence (third panel left), the mock host galaxy convergence (third panel centre) and the NFW subhalo convergence (third panel right). The bottom panel shows the reconstructed total convergence (bottom left), 
optimal reconstruction of the source's surface brightness (bottom centre) and the reconstructed added perturbation in convergence due to the subhalo (bottom right).

\begin{figure*}[htbp]
\centering
 \includegraphics[width=1\columnwidth]{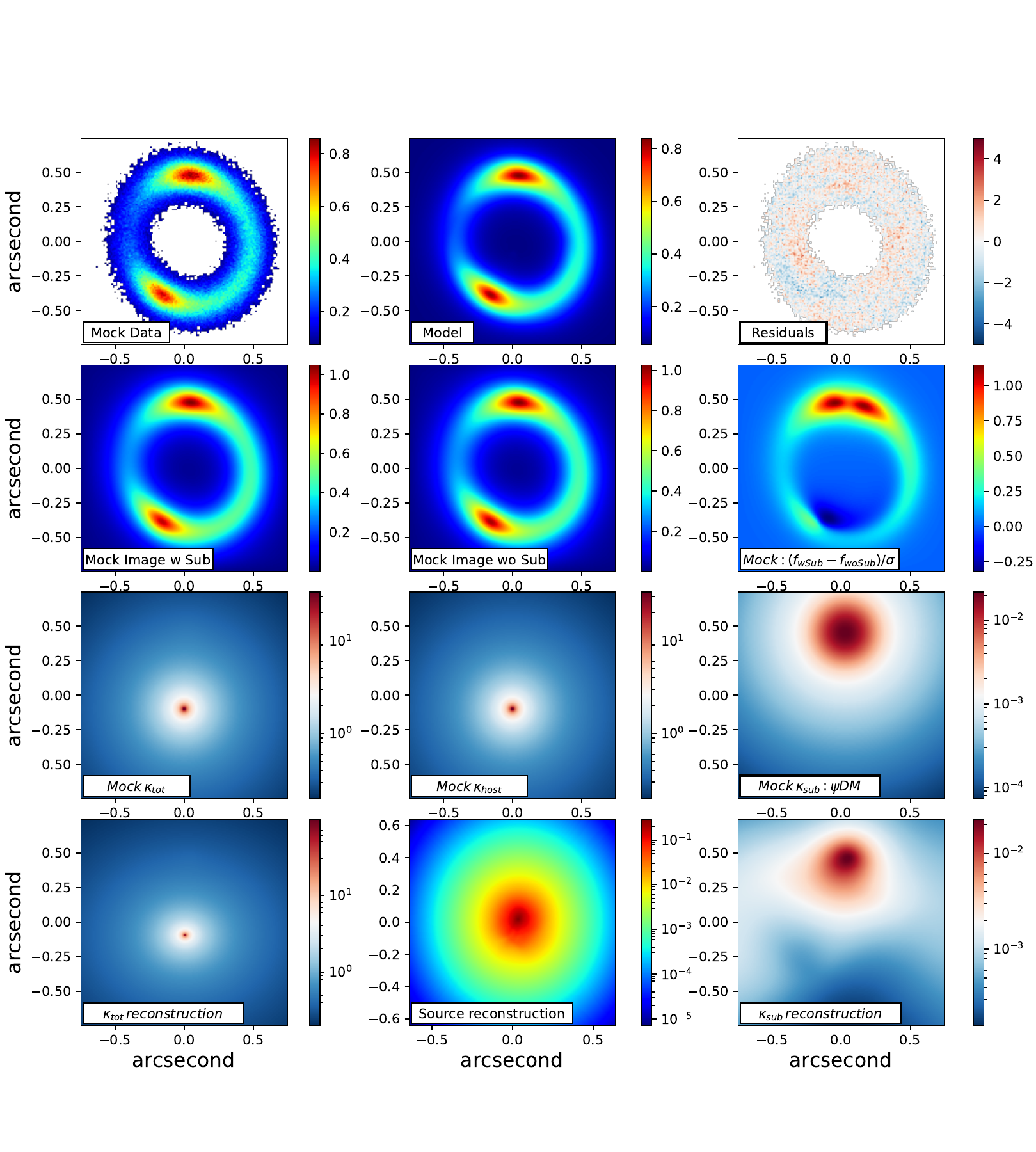}
 \caption{ The mock system with a $\psi$DM subhalo. }
 \label{fig:subhalo_mock_syserr_FDM}
\end{figure*}

For comparison, we also consider the simulation of $\psi$DM sub-halo and fit it. The simulated image is shown in Figure \ref{fig:subhalo_mock_syserr_FDM}. We consider a $\psi$DM subhalo with a core of $r_c\sim 0.6 \, kpc$, and has an NFW outskirt at a larger radius. Similarly, the subhalo mass we simulate is $M(r\leq 4\, kpc)=3\times 10^8\, M_{\odot}$. 

As shown in Figure~\ref{fig:subhalo_mock_syserr_FDM}. We fitted the mock data with \emph{LensCharm}. The top panel of Figure~\ref{fig:subhalo_mock_syserr_FDM} displays the mock data (top left), the reconstructed model (top centre), the residual (top right). The second panel shows the mock image with $\psi$DM subhalo (second panel left), the mock image without subhalo (second panel centre), the SNR of difference between mock images with and without subhalo (second panel right). The third panel displays the total mock lens convergence (third panel left), the mock host galaxy convergence (third panel centre) and the $\psi$DM subhalo convergence (third panel right). The bottom panel shows the reconstructed total convergence (bottom left), 
optimal reconstruction of the source's surface brightness (bottom centre) and the reconstructed added perturbation in convergence due to the subhalo (bottom right).

\begin{figure*}[htbp]
\centering
 \includegraphics[width=0.49\columnwidth]{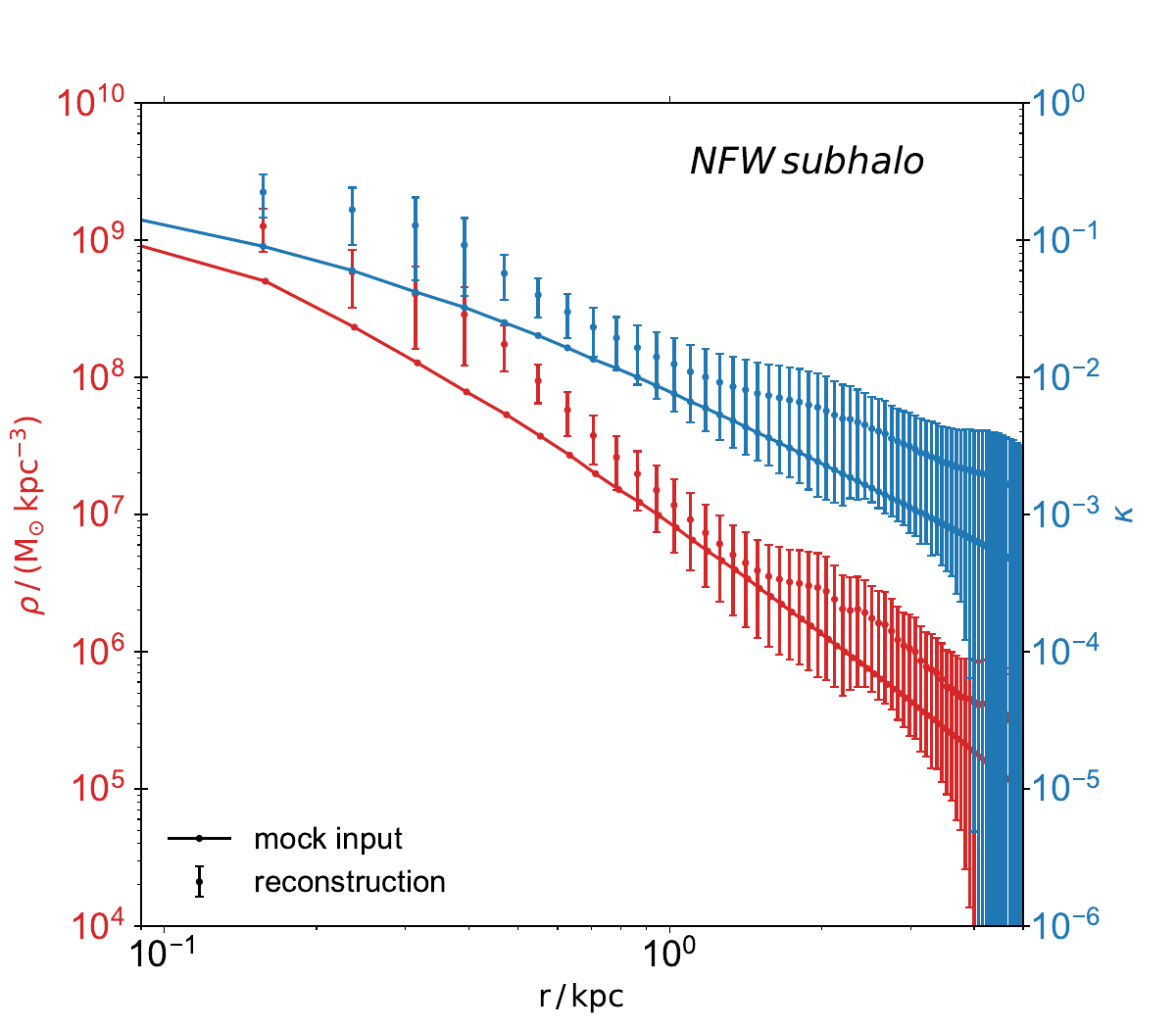}
 \includegraphics[width=0.49\columnwidth]{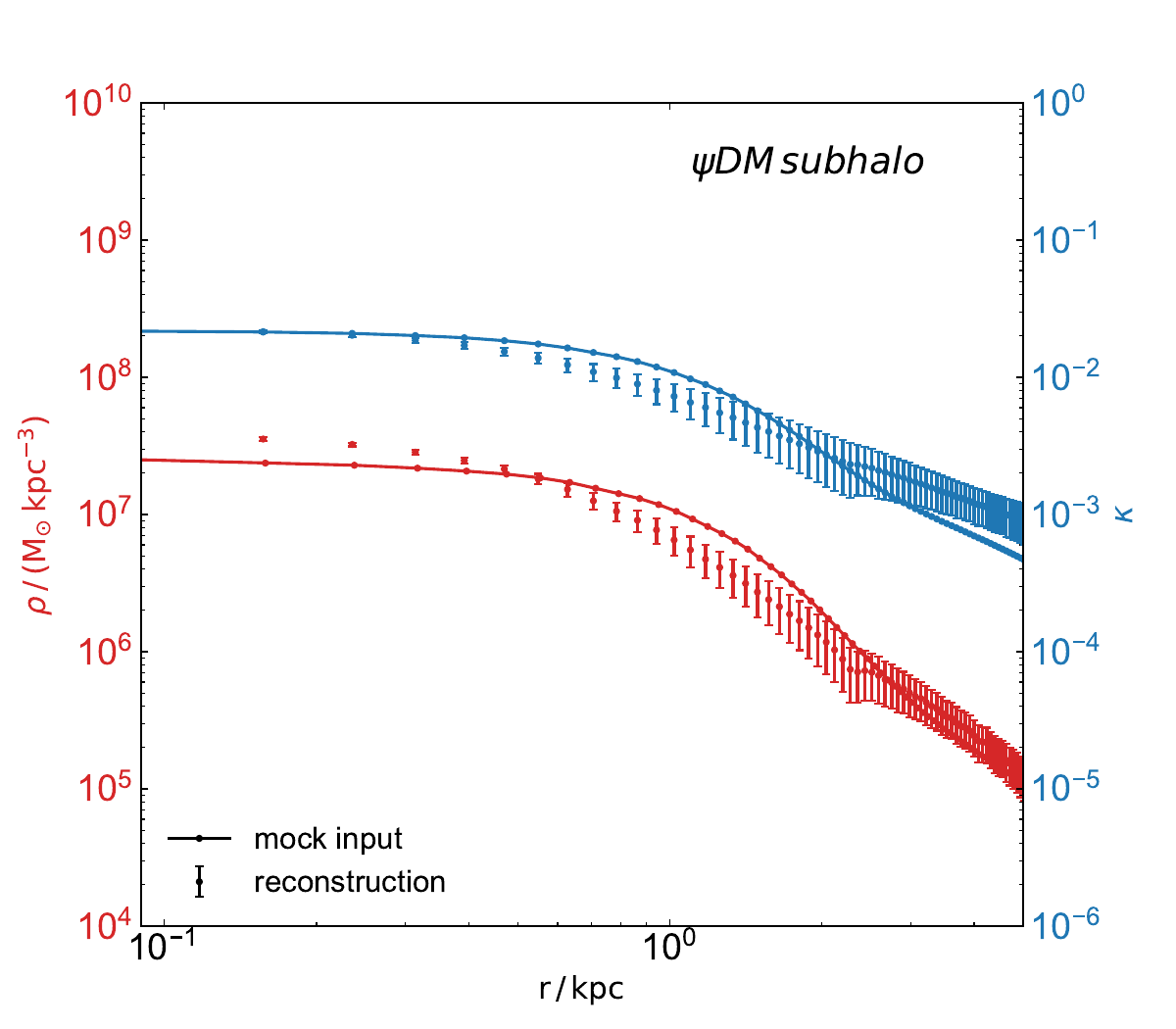}
 \caption{ The input and reconstruction of the mock system in NFW and $\psi$DM cases. The red profiles and data points with errorbars are the input subhalo 3D-density and reconstuction of density with \emph{LensCharm}. The blue profiles and data points are input convergence profiles and reconstructed profiles of convergence. Left Panel: The input and reconstructed density profile for a NFW subhalo.  Right Panel: The input and reconstructed density profile for a $\psi$DM subhalo. The corresponding image information is shown in Fig. \ref{fig:subhalo_mock_syserr} (NFW subhalo) and Fig. \ref{fig:subhalo_mock_syserr_FDM} ($\psi$DM subhalo).   }
 \label{fig:subhalo_mock_syserr_FDM_CDM}
\end{figure*}

After obtaining the best fit results, we compare them with the input values. As shown in Figure \ref{fig:subhalo_mock_syserr_FDM_CDM}, we find that the profiles of the two different dark matter halos can be reconstructed. In Figure \ref{fig:subhalo_mock_syserr_FDM_CDM}, we plot 3D density and 2D density of subhalo in red and blue, respectively. The solid line is the input value when we simulate the image. The data points are the result of our refactoring using \emph{LensCharm}, which are correspending to the $\kappa_{sub}$ reconstructions in the bottom panel right of Figure~\ref{fig:subhalo_mock_syserr} (NFW subhalo) and Figure \ref{fig:subhalo_mock_syserr_FDM} ($\psi$DM subhalo). 

With the best-fit non-parametric subhalo model of the mock data, we get the relative convergence deviations between the mock input of the NFW subhalo profile and the best-fit non-parametric subhalo profile.  Thus, we added the relative convergence error to the reconstructed JVAS B1938+666 subhalo density profile. The errors listed in Table~\ref{tab:data} and plotted in Figure~\ref{fig:result} are total errors including systematic errors from the above methods and statistic errors from the final best-fit convergence. In fitting different dark matter profile models, we also used the total errors.

\subsection{Reconstrcuted non-parametric dark matter density and mass}

The dark matter enclosed mass within radius $r$ is defined as  
\begin{equation}
M(r)=\int^r_0 4 \pi r^2 \rho(r) dr.
\label{eq:mass_rho_r}
\end{equation}
Here $\rho(r)$ is the three-dimensional density profile of the dark matter halo.

The surface density $\kappa$ is integrated line-of-sight density $\rho$:
\begin{equation}
\kappa (R) = \frac{1}{\Sigma_{cr}}\int^{+\infty}_{-\infty} \rho(\sqrt{R^2+z^2})  \, dz.
\label{eq:rho_to_kappa}
\end{equation}
 
Thus, the three-dimensional density profile can be calculated with the convergence by the Abel transform assuming the subhalo is symmetry spherically \citep{2007ApJ...655..135N,2024ApJ...970..143T,2025MNRAS.536.2672L}:
\begin{equation}
\rho(r) = -\frac{\Sigma_{cr}}{\pi} \int^{\infty}_{r} \frac{ \frac{d\kappa}{dR} }{\sqrt{R^2 - r^2}} dR ,
\label{eq:rho_kappa_r}
\end{equation}
where $\frac{d\kappa}{dR}$ is the derivative of $\kappa$, which can be calculated numerically: 
\begin{equation}
    \frac{d\kappa}{dR}|_{R_i} \approx \frac{\kappa (R_{i+1})-\kappa(R_{i-1})}{2\Delta R}.
\end{equation}

The model-independent density distribution and enclosed mass of the JVAS B1938+666 subhalo are listed in Tabel~\ref{tab:data}. The systematic errors and statistical errors from strong lens modelling of the system have been considered in the table. The five columns are 
radius, density $\rho$ at the radius, density error $\sigma_{\rho}$, enclosed mass $M(<r)$ in the inner region of $r$ and the corresponding error $\sigma_{M(<r)}$, respectively. Here we excluded the points outer than Einstein ring radius $r>6.4$ kpc because of the large errors and size limitation used in lens modelling.
\begin{longtable}{c|c|c|c|c}
    \hline
    \hline
    Radius & $\rho /10^6$ & $\sigma_{\rho} / 10^6$ & $M(<r) / 10^8$ & $\sigma_{M(<r)} / 10^8$ \\
    (kpc) & ($ \rm M_{\odot}/kpc^3 $) & ($  \rm M_{\odot}/kpc^3$)  & ($  \rm M_{\odot}$) & ($ \rm M_{\odot}$)  \\
    \hline
    \hline
0.079  &  25.285  &  5.123  &  0.001  &  0.0001 \\ 
0.158  &  23.843  &  4.831  &  0.004  &  0.001 \\ 
0.238  &  23.098  &  6.244  &  0.013  &  0.004 \\ 
0.317  &  22.061  &  6.187  &  0.031  &  0.009 \\ 
0.396  &  20.776  &  5.392  &  0.058  &  0.015 \\ 
0.475  &  19.257  &  4.064  &  0.096  &  0.02 \\ 
0.555  &  17.595  &  2.607  &  0.144  &  0.021 \\ 
0.634  &  15.963  &  1.407  &  0.203  &  0.018 \\ 
0.713  &  14.422  &  1.605  &  0.271  &  0.03 \\ 
0.792  &  12.984  &  2.539  &  0.349  &  0.068 \\ 
0.872  &  11.684  &  3.572  &  0.433  &  0.133 \\ 
0.951  &  10.485  &  4.488  &  0.525  &  0.225 \\ 
1.03  &  9.284  &  5.045  &  0.621  &  0.338 \\ 
1.109  &  8.272  &  5.615  &  0.721  &  0.489 \\ 
1.189  &  7.15  &  5.559  &  0.822  &  0.639 \\ 
1.268  &  6.314  &  5.707  &  0.923  &  0.834 \\ 
1.347  &  5.53  &  5.513  &  1.023  &  1.02 \\ 
1.426  &  4.945  &  5.497  &  1.123  &  1.249 \\ 
1.506  &  4.685  &  5.658  &  1.226  &  1.481 \\ 
1.585  &  3.748  &  4.782  &  1.326  &  1.691 \\ 
1.664  &  3.259  &  4.451  &  1.417  &  1.936 \\ 
1.743  &  3.21  &  4.602  &  1.511  &  2.166 \\ 
1.823  &  2.896  &  4.319  &  1.607  &  2.397 \\ 
1.902  &  2.611  &  4.03  &  1.702  &  2.628 \\ 
1.981  &  2.131  &  3.368  &  1.791  &  2.83 \\ 
2.06  &  1.733  &  2.808  &  1.869  &  3.028 \\ 
2.139  &  1.677  &  2.774  &  1.944  &  3.216 \\ 
2.219  &  1.602  &  2.687  &  2.021  &  3.389 \\ 
2.298  &  1.314  &  2.213  &  2.095  &  3.529 \\ 
2.377  &  1.056  &  1.793  &  2.159  &  3.664 \\ 
2.456  &  1.008  &  1.723  &  2.219  &  3.794 \\ 
2.536  &  0.97  &  1.659  &  2.28  &  3.902 \\ 
2.615  &  0.875  &  1.495  &  2.341  &  3.999 \\ 
2.694  &  0.758  &  1.285  &  2.398  &  4.066 \\ 
2.773  &  0.653  &  1.077  &  2.451  &  4.042 \\ 
2.853  &  0.605  &  0.988  &  2.5  &  4.084 \\ 
2.932  &  0.556  &  0.891  &  2.548  &  4.082 \\ 
3.011  &  0.503  &  0.798  &  2.595  &  4.121 \\ 
3.09  &  0.456  &  0.711  &  2.639  &  4.109 \\ 
3.17  &  0.411  &  0.626  &  2.682  &  4.08 \\ 
3.249  &  0.378  &  0.571  &  2.722  &  4.115 \\ 
3.328  &  0.346  &  0.513  &  2.761  &  4.086 \\ 
3.407  &  0.317  &  0.463  &  2.798  &  4.085 \\ 
3.487  &  0.291  &  0.416  &  2.834  &  4.047 \\ 
3.566  &  0.27  &  0.378  &  2.869  &  4.013 \\ 
3.645  &  0.25  &  0.345  &  2.903  &  4.012 \\ 
3.724  &  0.229  &  0.31  &  2.935  &  3.965 \\ 
3.804  &  0.212  &  0.284  &  2.966  &  3.962 \\ 
3.883  &  0.197  &  0.257  &  2.996  &  3.908 \\ 
3.962  &  0.184  &  0.235  &  3.025  &  3.862 \\ 
4.041  &  0.172  &  0.216  &  3.054  &  3.831 \\ 
4.12  &  0.161  &  0.198  &  3.082  &  3.792 \\ 
4.2  &  0.15  &  0.183  &  3.108  &  3.774 \\ 
4.279  &  0.141  &  0.169  &  3.134  &  3.749 \\ 
4.358  &  0.131  &  0.154  &  3.16  &  3.703 \\ 
4.437  &  0.124  &  0.143  &  3.184  &  3.674 \\ 
4.517  &  0.118  &  0.134  &  3.208  &  3.638 \\ 
4.596  &  0.109  &  0.122  &  3.232  &  3.599 \\
    \hline
    \caption{ The dark matter distribution of the subhalo in strong-lensing system JVAS JVAS B1938+666. This is detected with the nonparametric strong lensing convergence perturbation method with \href{https://gitlab.mpcdf.mpg.de/ift/lenscharm}{\emph{LensCharm}}. The values in the five columns are: radius, density $\rho$ at the radius, density error $\sigma_{\rho}$, enclosed mass $M(<r)$ in the inner region of $r$ and the corresponding error $\sigma_{M(<r)}$. }
\label{tab:data}
\end{longtable}

\section{Posterior of dark matter profile models}
\label{app:posterior}

We emploied the MCMC code emcee \citep{2013PASP..125..306F} to give the posterior of the dark matter density profile parameters fitted with our reconstructed non-parametric subhalo profile of JVAS B1938+666. The prior distributions of the parameters are listed in Tabel~\ref{tab:prior}. 

\begin{figure*}[ht!]
\centering
 \includegraphics[width=0.32\columnwidth]{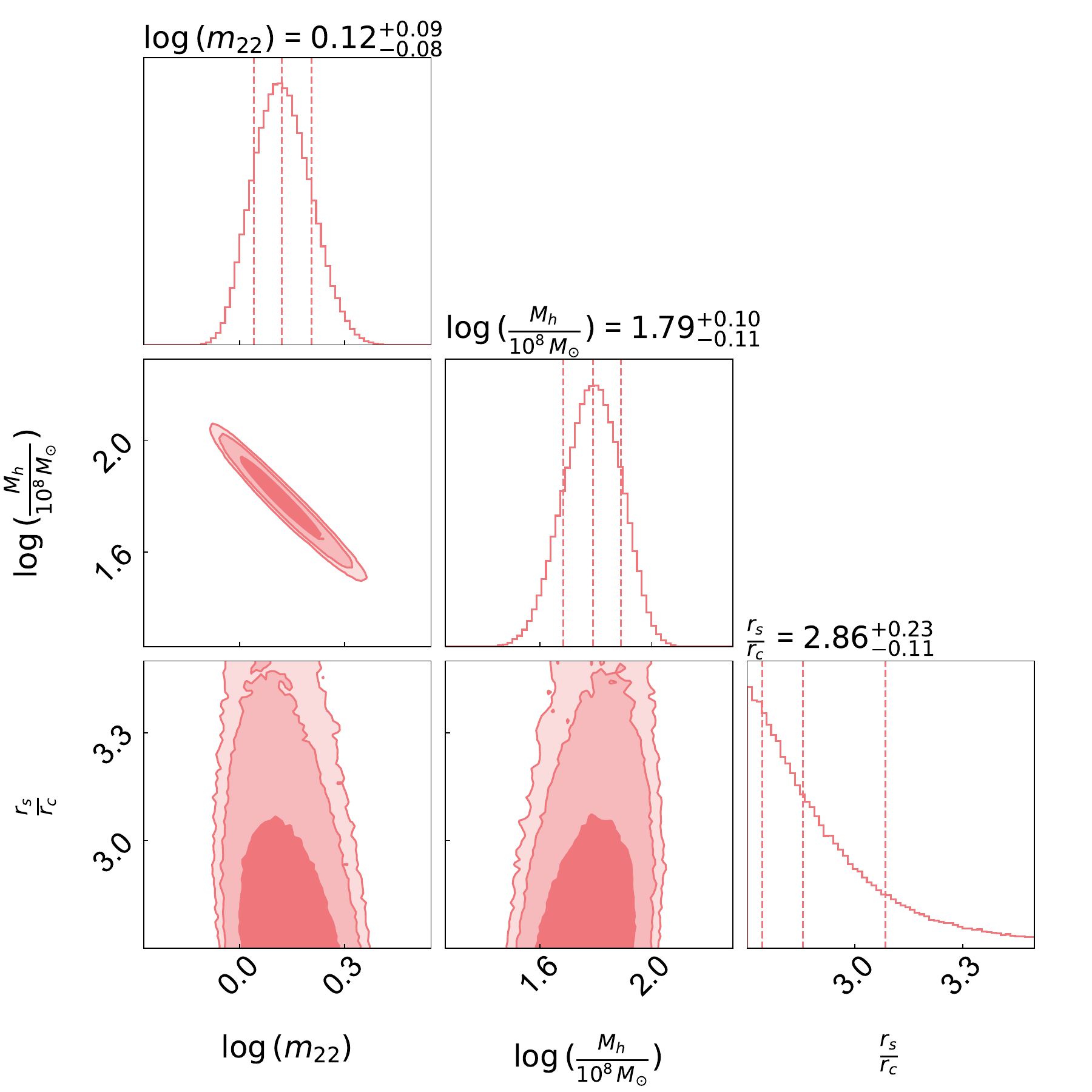}
 \includegraphics[width=0.32\columnwidth]{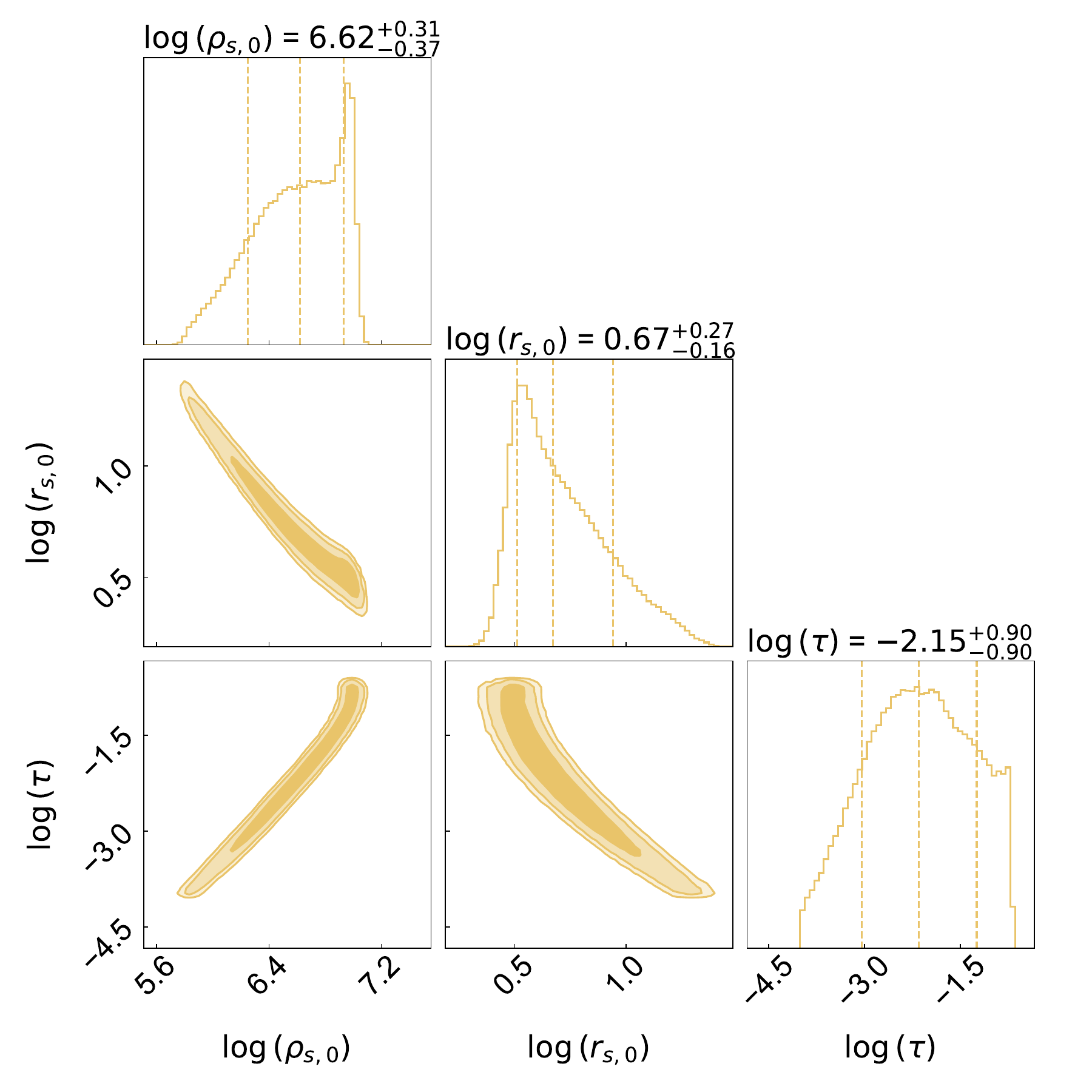}
 \includegraphics[width=0.32\columnwidth]{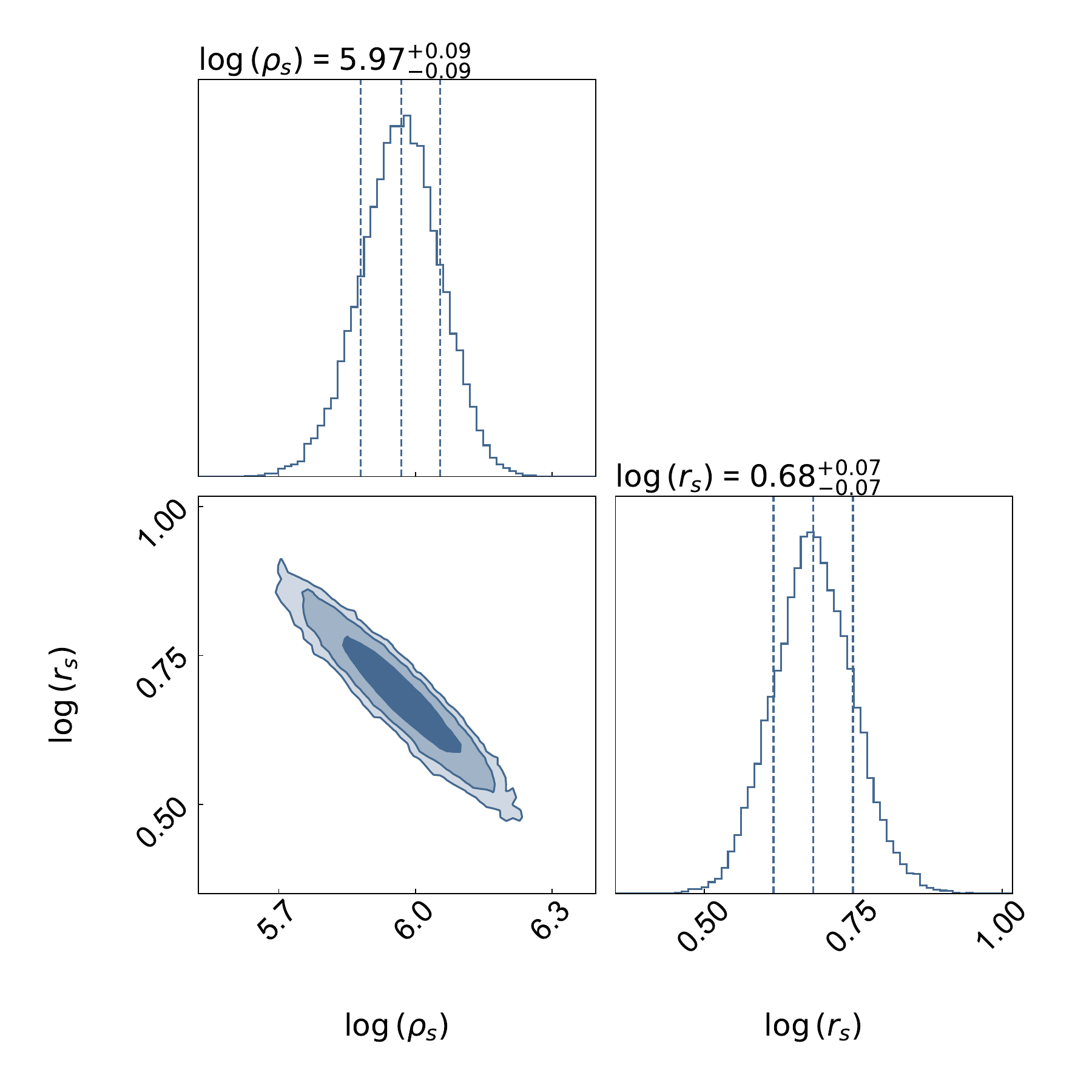}
 
 \caption{The posterior distribution of the dark matter profile models from fitting the mass distribution of JVAS JVAS B1938+666. The vertical dashed lines in histogram plots show percentage values in 16\%, 50\% and 84\%. Left panel: The posterior of parameters of the $\psi$DM model. Central panel: The posterior of the SIDM profile parameters. Right panel: The posterior of the NFW profile. }
 \label{fig:posterior}
\end{figure*}

Figure~\ref{fig:posterior} shows the posteriors of the three dark matter profile models fitted with the reconstructed non-parametric subhalo density profile. The vertical dashed lines in histogram plots show percentage values in 16\%, 50\% and 84\%. The left, central and right panels show the posteriors of parameters of the $\psi$DM, SIDM and the NFW profiles. 

\bibliography{sample631}{}
\bibliographystyle{aasjournal}


\end{CJK*}

\end{document}